\documentclass[num-refs]{wiley-article}

\usepackage{siunitx}
\usepackage{microtype}
\usepackage{multicol}
\usepackage{float}
\usepackage{tabularx}

\papertype{Pre-print article}
\paperfield{Draft}

\title{Analysis of the SARS-CoV-2 outbreak in \\ Rio Grande do Sul / Brazil}

\abbrevs{RS, Rio Grande do Sul.}

\author[1]{Christian S. Perone}
\affil[1]{BSc. Computer Science, Universidade de Passo Fundo/RS (UPF), Brazil. MSc. Biomedical Engineering, Polytechnique Montreal/UdeM, Canada.}
\corraddress{\url{http://blog.christianperone.com}}
\corremail{christian.perone@gmail.com}

\fundinginfo{This is a draft and is subject to correction, it was \textbf{not peer reviewed} and I have no intention in submitting it to a journal or conference. Take it as is and be critical. The main goal of this draft is to try to provide a better characterization of the outbreak in Rio Grande do Sul (RS) and especially Porto Alegre/RS. If you see any issues in this analysis, please contact the author.}

\runningauthor{Analysis of the SARS-CoV-2 outbreak in Rio Grande do Sul/Brazil -- Christian S. Perone}

\begin{document}

\begin{frontmatter}
\maketitle

\begin{abstract}
This article contains a series of analyses done for the SARS-CoV-2 outbreak in Rio Grande do Sul (RS) in the south of Brazil. These analyses are focused on the high-incidence cities such as the state capital Porto Alegre and at the state level. We provide methodological details and estimates for the effective reproduction number $R_t$, a joint analysis of the mobility data together with the estimated $R_t$ as well as ICU simulations and ICU LoS (length of stay) estimation for hospitalizations in Porto Alegre/RS.

% Please include a maximum of seven keywords
\end{abstract}
\end{frontmatter}

\section{Introduction}
The first confirmed reported case of SARS-CoV-2 in Rio Grande do Sul (RS), south of Brazil, was on March 10th, 2020. The subject was a 60 years old man living in Campo Bom/RS, a city 50km far from the capital of the state: Porto Alegre. The first confirmed patient had mild symptoms and didn't require hospitalization. On the confirmation date of the first case, on March 10th, 190 subjects were already reported as suspected cases \cite{ses_2020}, with 103 discarded and 86 in investigation.

At the time of writing of this paper, data from Secretaria Estadual da Saúde (SES-RS) shows that the state has 46.710 confirmed cases on July 17th, with 1.174 deaths and an adult intensive care unit (ICU) occupancy of 76\% on the entire state with 580 confirmed cases at the ICU units \cite{ses_2020_2}.

\section{Effective reproduction number $R_t$ estimation}
The effective reproduction number, often called $R(t)$, $R_t$ or $R_e$ is the expected number of new infections that are caused by a primary case, on a context where not the entire population are susceptible, as opposed to the basic reproduction number $R0$ \cite{Delamater2019}, where the entire population is assumed to be susceptible.

Since April 12th, we made available \footnote{At the website: \url{https://perone.github.io/covid19analysis/}} the $R_t$ estimates for all states in Brazil, including Rio Grande do Sul. Choosing a method to estimate the $R(t)$ is a challenging task. Many methods with different characteristics were developed in the past \cite{Cori2013,Wallinga2004,Bettencourt2008}, and deciding which method to use heavily depends on the context and the properties that the context requires.

We opted to use the method from \cite{Cori2013}, which uses the serial interval (SI) distribution -- the time between successive clinical onsets in a chain of transmission --, which is a quantity that is easier to observe than the generation time, since times of infection are rarely observed. The method from \cite{Cori2013} is based on the relationship of the incidence $I_t$ at time $t$, the infectivity profile $w_s$ that is approximated by the distribution of the serial interval and the $R_t$, as given by the following definition:

\begin{equation} \label{eq:1}
\mathbb{E}[I_t] = R_t \sum_{s=1}^{t} I_{t-s}w_s
\end{equation}

The relationship shown in the Equation \ref{eq:1} tells us that we are weighting the incidence by symptom onset using the infectivity profile, and then multiplying it by the $R_t$. To reduce the variance of estimating the $R_t$ from the equation above, the authors calculate estimates over longer time windows (assuming that the instantaneous reproduction number is constant within that window). In our case, we used a 1 week period, but longer periods can be used to reduce variance, but might also hide some dynamics of the $R_t$ or lead to lagged and inaccurate $R_t$ estimates.

Another advantage of the method from \cite{Cori2013}, is that the posterior can be easily calculated when assuming a Gamma prior distribution on the $R_t$. The method also allows us to take into consideration the uncertainty around the serial interval distribution, a topic of constant debate \cite{Jr.2020,Ashcroft2020}, and therefore with high uncertainty associated.

Another important property that was also one of the key reasons to select the method was that the method is reasonably robust to under-reporting as shown in the sensitivity analysis on various scenarios made by the authors. As this is a known issue in many states of Brazil, this method poses a strong advantage as long as the testing bias is constant, which is often not true at the beginning of the outbreak, but important as testing capacity tends to stabilize afterward.

Besides the aforementioned reasons to select the method \cite{Cori2013}, recently, a study found that the method \cite{Bettencourt2008} systematically underestimates $R_t$ when the true value is substantially higher than one, while also being biased as transmission rates shift and recommended the method from \cite{Cori2013}.

\subsection{Adjusting for right-censoring}
\begin{figure}[h]
	\centering
	\includegraphics[width=\textwidth]{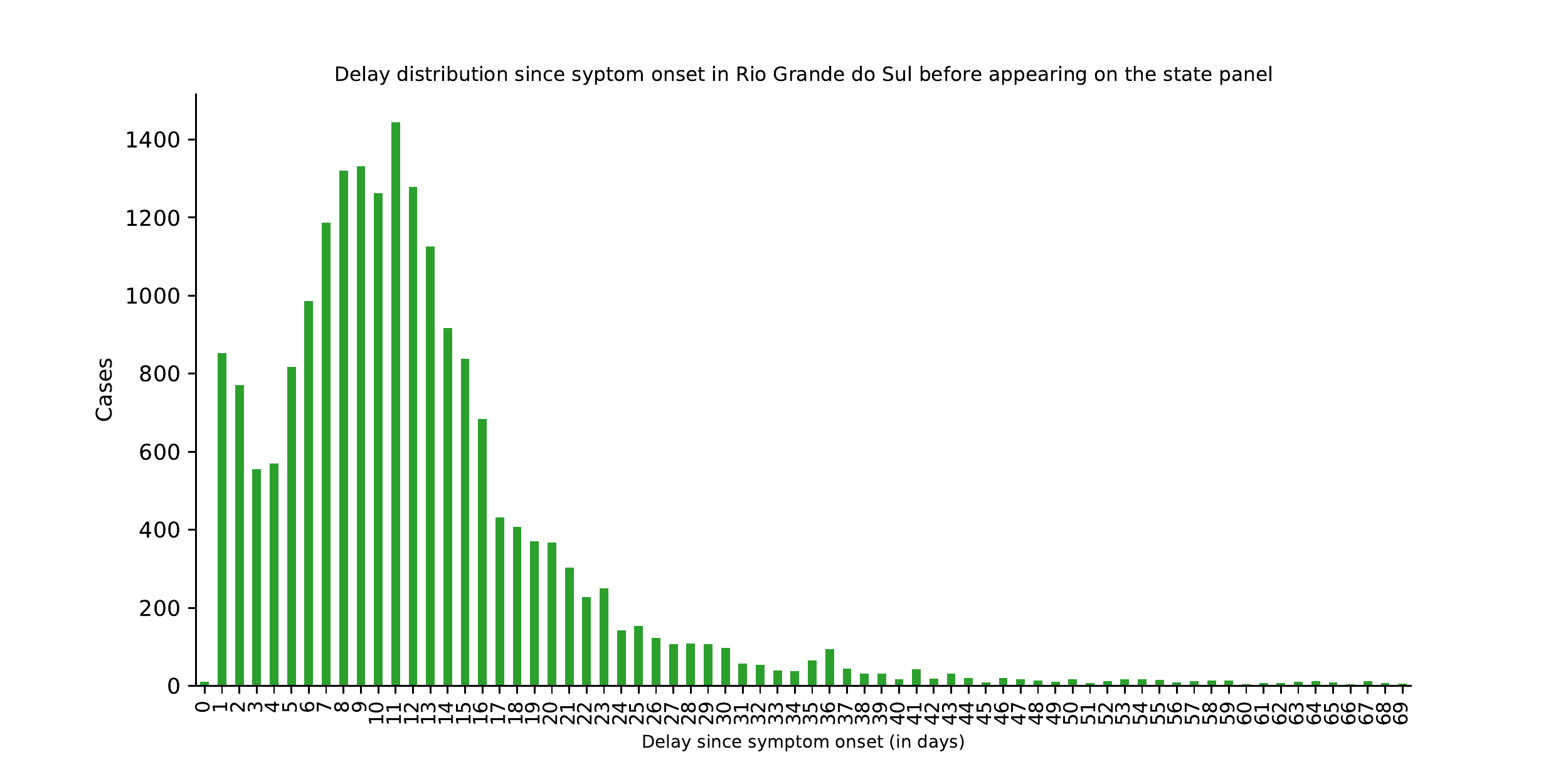}
	\caption{Empirical distribution of the delay (in days) between the symptom onset and for the case to appear in the state main panel where data is made available. Data provided by SES-RS from Rio Grande do Sul (RS) / Brazil, using 17 days ranging from June 13th to July 17th. Distribution was truncated to represent only 70 values.}
	\label{fig:delay_rs}
\end{figure}

It is well-known that the SARS-CoV-2 outbreak in Brazil posed a lot of new challenges regarding the testing procedures that required strategical coordination among many different entities from different cities. This pressure caused major delays in many states of Brazil and by looking at the Figure \ref{fig:delay_rs}, where we show the delay distribution since symptom onset until cases appear in the main panel of the SES-RS, there is strong evidence of major delays taking place in the state procedures. Cases from Porto Alegre/RS, the state capital, also suffered major delays and on a single day, more than 1600 cases were added on a batch that had cases with more than one month of delay since symptom onset.

Given this context, it is even more important to deal with that truncation properly. Many statistical methods exist to deal with right-censoring, including parametric distribution estimation or non-parametric methods. Most of the time, these methods rely on the estimation of an Exponential, Gamma ou Weibull distribution and then uses the (flipped) cumulative density function (CDF) to correct for the delays.

\begin{figure}[h]
	\centering
	\includegraphics[width=\textwidth]{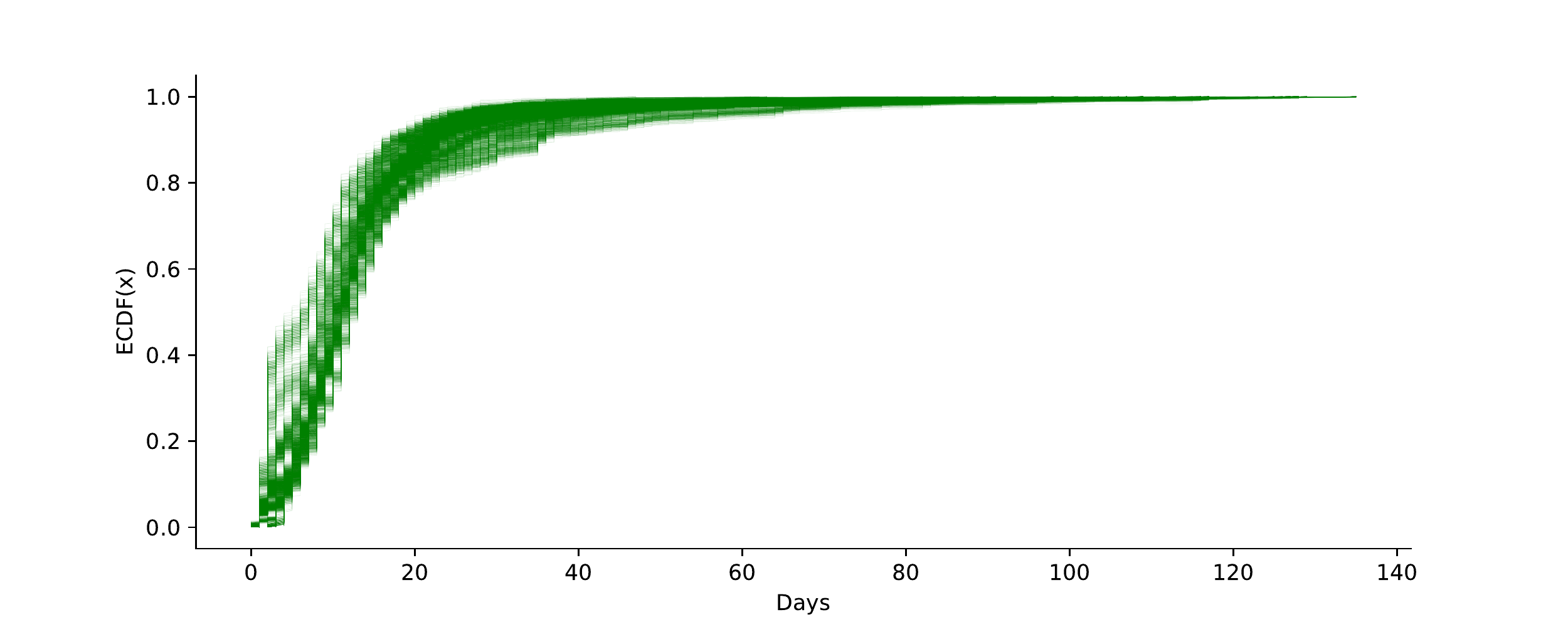}
	\caption{Bootstrap samples from the ECDF showing the uncertainty around the delay distribution for Rio Grande do Sul (RS) / Brazil.}
	\label{fig:ecdf_rs}
\end{figure}

Parametric methods are often sample efficient and work well on the small data regime. However, they fail to model the complexity of distributions under model misspecification, especially when there are multiple modes due to complex and dynamic processes. Since we have a reasonable amount of historical data for state level analysis, a better approach is to use the empirical CDF (ECDF) and then do bootstrapping to accommodate for the uncertainty around the observed delays, as shown in Figure \ref{fig:ecdf_rs}. The ECDF in our case is constructed using the delay distribution by aggregating the new symptom onset dates at each new day, and using only a few recent weeks of data to capture the most recent delay patterns.

\begin{figure}[h]
	\centering
	\includegraphics[width=\textwidth]{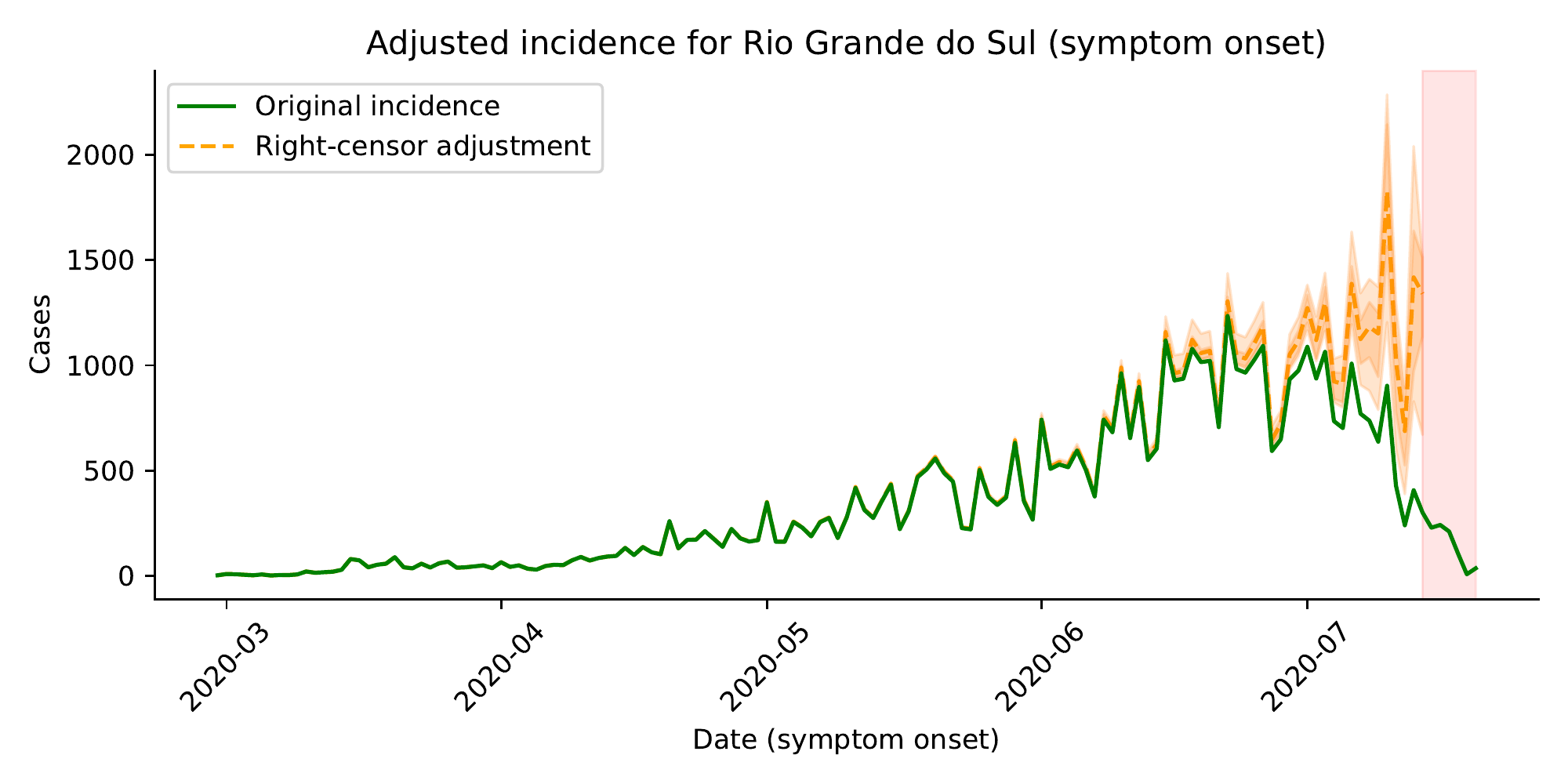}
	\caption{Result of the right-censoring adjustment with a HDI interval of 0.7 and 0.95 respectively shown as orange bands. The red region are the discarded samples.}
	\label{fig:censor_adjustment}
\end{figure}

The ECDF is then used to parametrize a Negative Binomial distribution as in \cite{Abbott2020}:

\begin{equation}
o_{t-j}^{*} \sim \text{NegBin}(n=o_{t-k}+1, p=F(t-j;\theta_i))
\end{equation}

where $F(t–j;\theta_i))$, differently than in \cite{Abbott2020}, is the ECDF for the bootstrapped delay distribution ($\theta_i$ sample). This estimation will give the proportion of onset cases from $t–j$ days ago that are expected to appear over the $j$ days from that time until the present on the SES-RS panel. The $t$, as in \cite{Abbott2020}, is the last date on which cases were reported, so the number of onsets $o_{t–j}$ on day $t-j$ is taken as a result of a Bernoulli trial from the total true number of cases with clinical onsets on day $t–j$. Like in \cite{Abbott2020} we also discard some days at the end (6 days in our case) and allow upscaling up to a limit of 5 times the reported cases for the day, as this approach also introduces an inevitable estimation bias. An example of this adjustment result can be seen in Figure \ref{fig:censor_adjustment}.

In order to propagate the right-censoring adjusment uncertainty into the $R_t$ estimation procedure, we sample 1000 samples from the bootstraped delay distribution described in the previous section, and then fit 1000 models using EpiEstim \cite{Cori2019} and a serial interval with $\mu = 4.7 (3.7 - 6.0)$ and $\sigma = 2.9 (1.9 - 4.9)$ \cite{Nishiura2020}. After estimating 1000 models, we sample from the assumed Gamma posterior of each model to build the final posterior and calculate the $R_t$ statistics such as mean, median, and highest density interval (HDI) for each day. Propagation of right-censoring adjustment uncertainty affects mostly in the most recent days of the distribution.

\subsection{Results of the $R_t$ estimation}
\begin{figure}[ht]
	\centering
	\includegraphics[width=16cm]{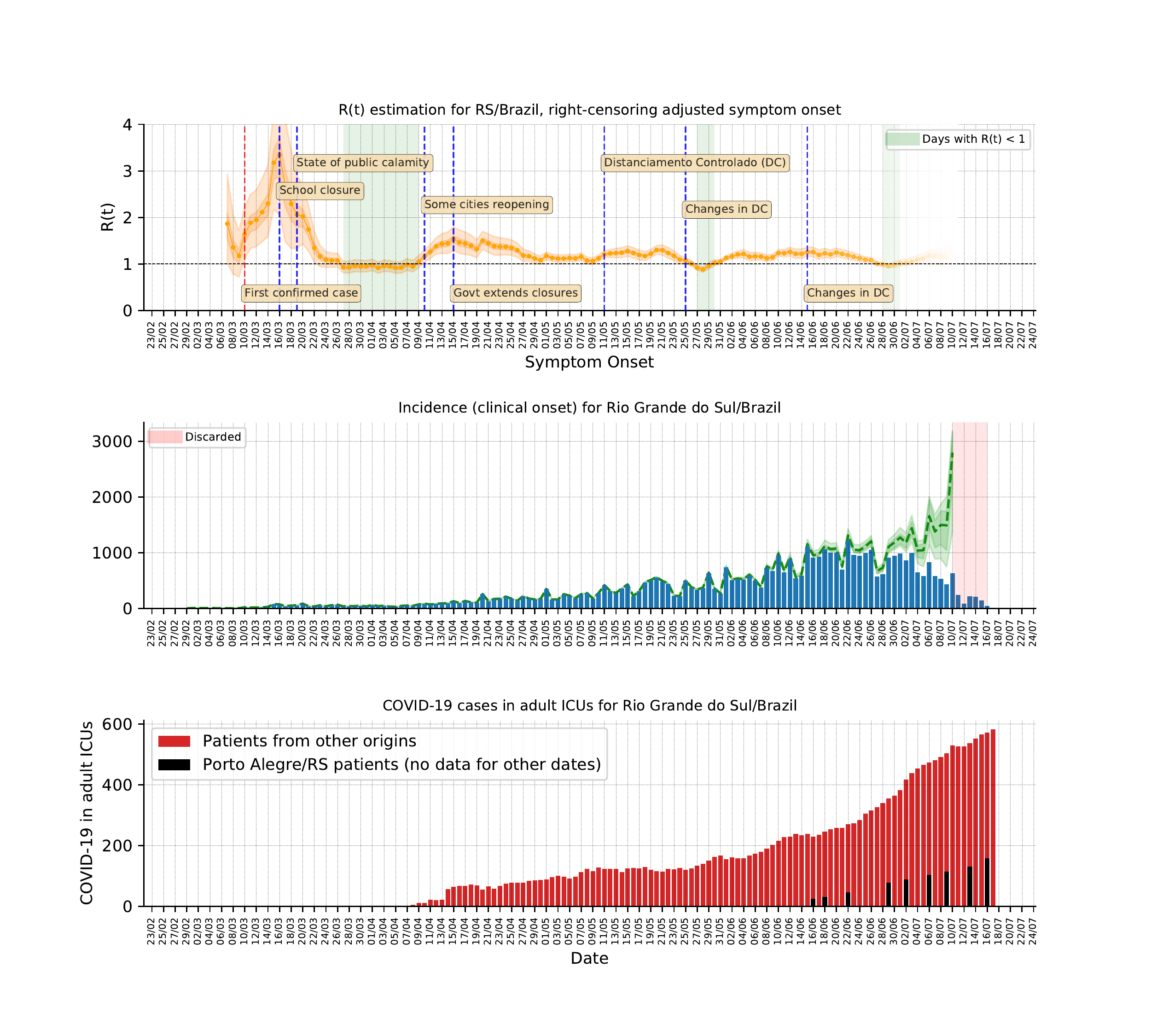}
	\caption{This plot shows the $R_t$ estimation for Rio Grande do Su (RS) / Brazil. \textbf{Top panel}: in this panel, you can see the $R_t$ estimate with orange bands representing the .5 and .975 posterior HDI, the center line is the median. The green vertical bands represent when the median $R_t$ is below 1.0. Labels denote some important events during the outbreak, such as revisions of the social distancing practices or closures of schools, etc. Transparency in the most recent days is used to show uncertainty due to the right-censor adjustment. \textbf{Middle panel}: in the middle panel you can see the incidence by symptom onset in blue and the green dashed line with the bands represent the right-censoring adjustment. The red vertical region represents discarded days as described in the methodology section. \textbf{Bottom panel}: in the bottom panel you can see adult ICU occupancy with confirmed COVID-19 cases. The black bards represent the ICU occupancy by Porto Alegre, the state capital.}
	\label{fig:rs_restim}
\end{figure}

In Figure \ref{fig:rs_restim}, we show the results of the $R_t$ estimation for the state of Rio Grande do Sul. These estimates used official data provided by SES-RS\footnote{Available at: \url{http://ti.saude.rs.gov.br/covid19/}} until July 17th, 2020.

As we can see in the Figure \ref{fig:rs_restim}, we had a median $R_t$ below 1.0 for a longer period only at the beginning of the outbreak when the interventions were potentially able to achieve its goals (but with the uncertainty interval above 1.0), after that period, coincidentally when interventions started to be relaxed (with local media entities reporting many cities reopening), the $R_t$ kept oscillating above 1.0 for a long period. We can also see at the bottom panel that ICU occupancy also started to grow, showing a very worrying and sustained trend that can cause a collapse of the health system as seen in other places \cite{Grasselli2020}.

\begin{figure}[h]
	\centering
	\includegraphics[width=\textwidth]{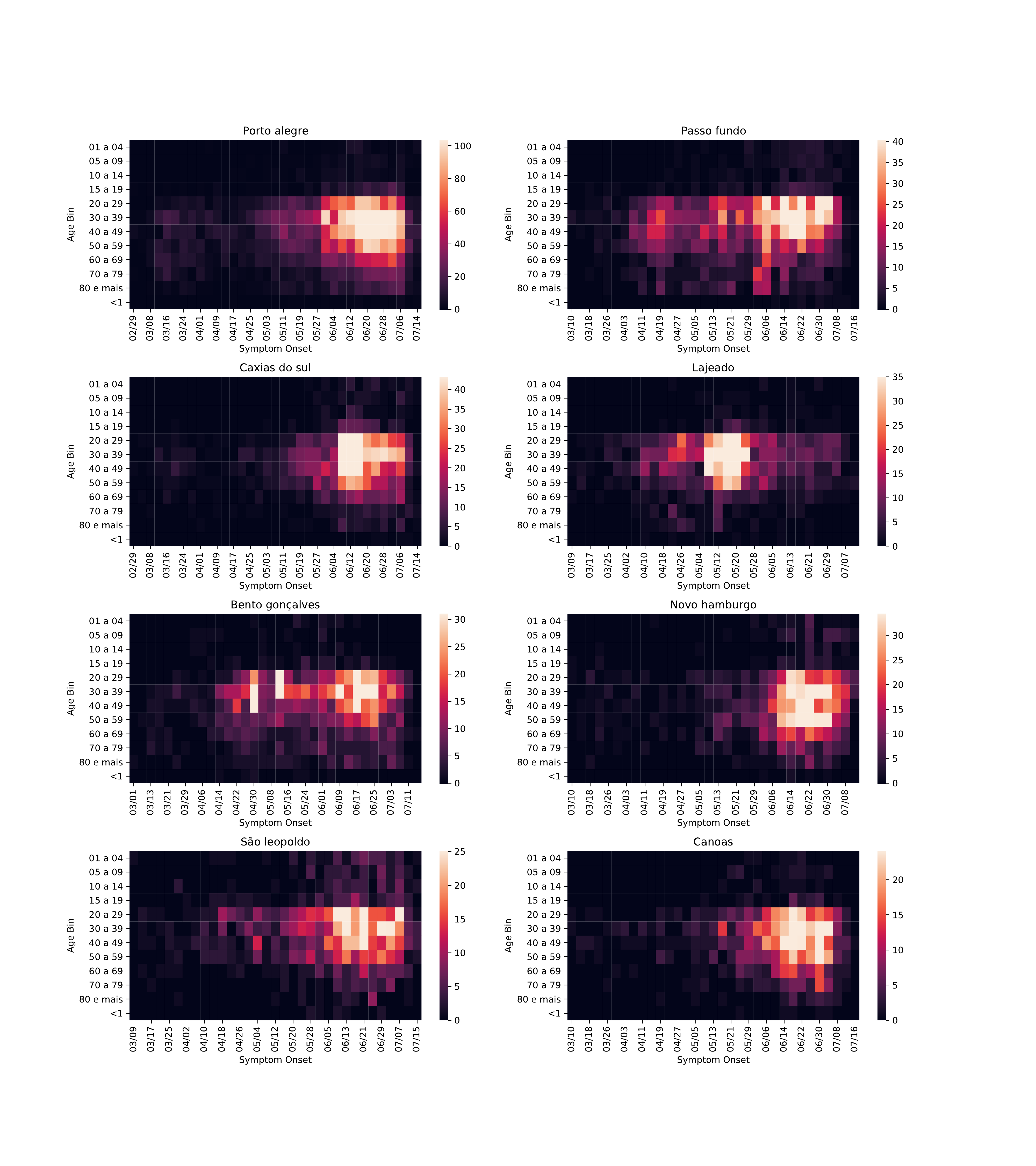}
	\caption{Heatmap of age distribution through time for confirmed COVID-19 cases on 8 cities with the most case incidence in Rio Grande do Sul/Brazil. Each square uses a 4-day aggregation period.}
	\label{fig:rs_agemap}
\end{figure}

In Figure \ref{fig:rs_agemap} we also show a heatmap of confirmed cases for 8 cities in RS with the most number of cases in the state. A pattern that was also seen in other places is shown, where the infections usually start at a young cohort before spreading to the elderly. Note that we did not adjust for right-censoring in this figure, therefore for all cities, it appears that there is a reduction of cases in the last days, but is usually an artifact of the delay from symptom onset to confirmation and reporting.

\subsection{$R_t$ and mobility data trajectories}
Many providers made available mobility data during the outbreak, one example is the COVID-19 Community Mobility Reports \cite{Google2020} that was made available early in the outbreak by Google. The dataset contains anonymized data (using techniques such as differential privacy) that prevents the identification of any individual person. These reports contain the movement trends by region, across different categories of places, and compared to a pre-outbreak baseline taken from a 5-week period between Jan 3rd and Feb 6th, 2020.

\begin{figure}[h]
	\centering
	\includegraphics[width=\textwidth]{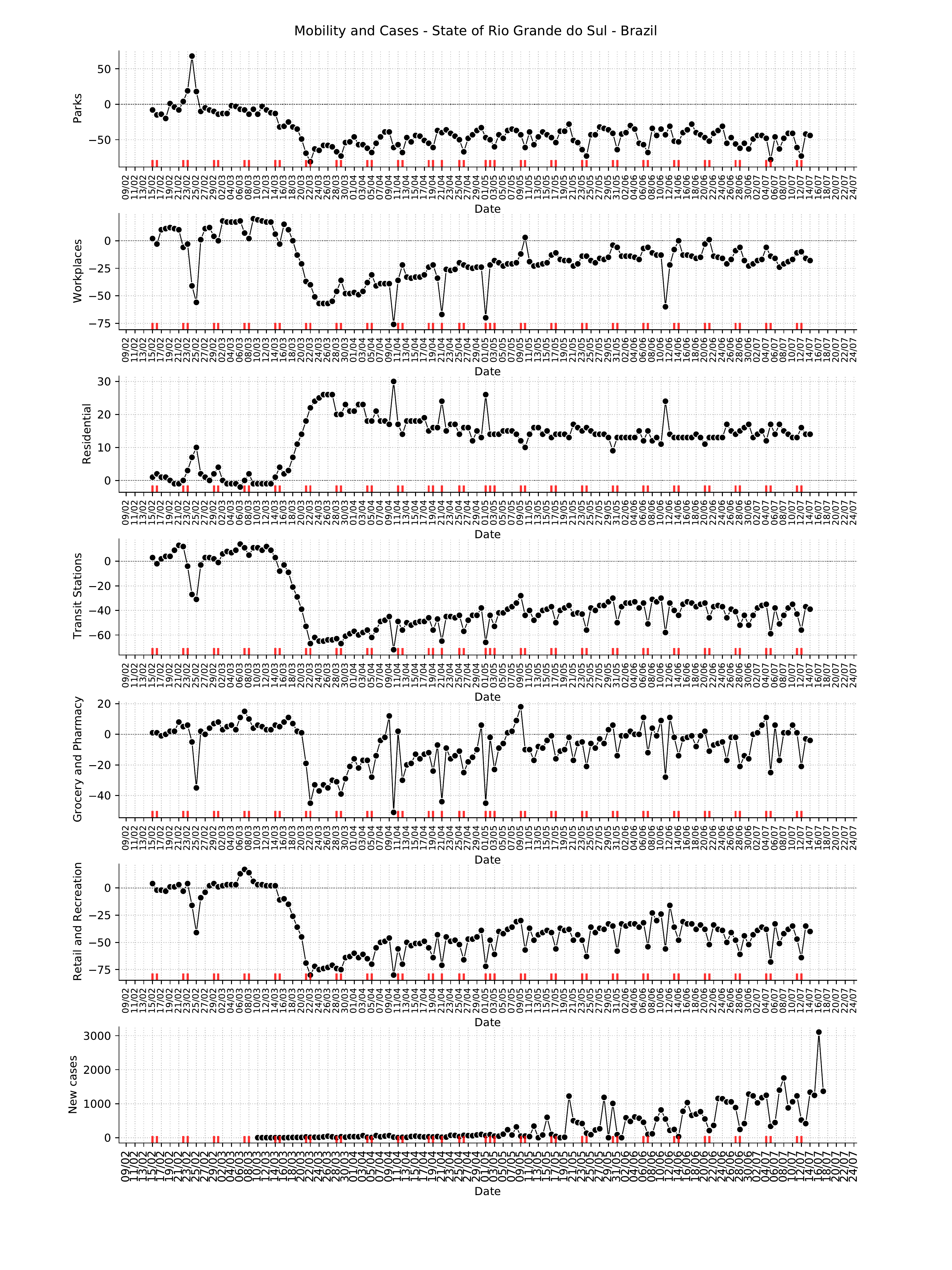}
	\caption{Data from the COVID-19 Community Mobility Reports \cite{Google2020} by Google for Rio Grande do Sul / Brazil. This report was collected on July 19th but the reports are provided lagged a few days by Google. In the last panel, we have the incidence by confirmation date, this data was collected from the Brazilian Ministry of Health (after aggregating SES-RS data). Red markers at the x-axis represent holidays or weekends in Brazil.}
	\label{fig:rs_mobility}
\end{figure}

The Community Mobility Reports show movement trends by region, across different categories of places: parks, workplaces, residential, transit stations, grocery and pharmacy, and retail and recreation. In Figure \ref{fig:rs_mobility} we can see this data for Rio Grande do Sul, separated by each different place category together with a incidence panel of confirmed cases at the bottom of the figure.

\begin{figure}[ht]
	\centering
	\includegraphics[width=14cm]{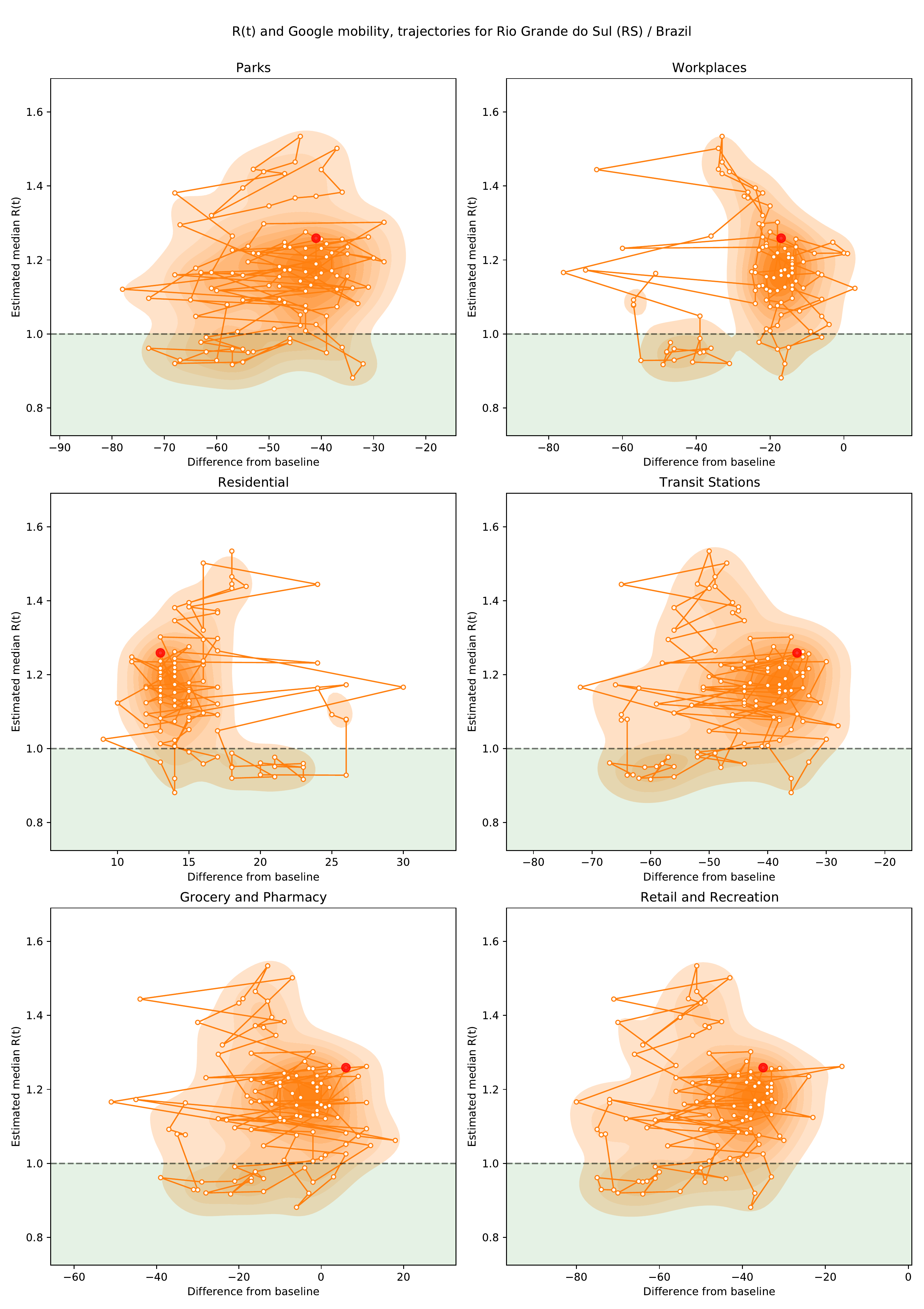}
	\caption{$R_t$ estimation trajectories with Google mobility data. Each panel of this plot shows different categories of mobility plotted together with the $R_t$ estimation of each date from March 23rd until July 10th. The red dot represents the last date. The green region represents the region where $R_t$ is below 1.0. The orange lines represent the joint point for the $R_t$ and the mobility difference from the pre-outbreak baseline. The orange shades represent the KDE estimation of the final trajectory distribution. Animation provided at: \protect\url{https://bit.ly/399On2N}.}
	\label{fig:mob_restim_rs}
\end{figure}

Adding a new dimension to the $R_t$ such as the mentioned mobility data can help visualize the dynamics of the $R_t$ and mobility at the same time. In Figure \ref{fig:mob_restim_rs} we show these joint trajectories of mobility and the estimated $R_t$ for Rio Grande do Sul. The density distribution at the background was estimated using a Gaussian kernel. As we can see in the figure, most of the density mass is concentrated above the median $R_t$ of 1.0. 

If we look at the \emph{workplaces} panel of Figure \ref{fig:mob_restim_rs} we can see a very clear separated density cluster representing the early moment at the outbreak when mobility was very low and with a respective $R_t$ below 1.0.

\section{$R_t$ estimation for Porto Alegre / RS}
\begin{figure}[h]
	\centering
	\includegraphics[width=16cm]{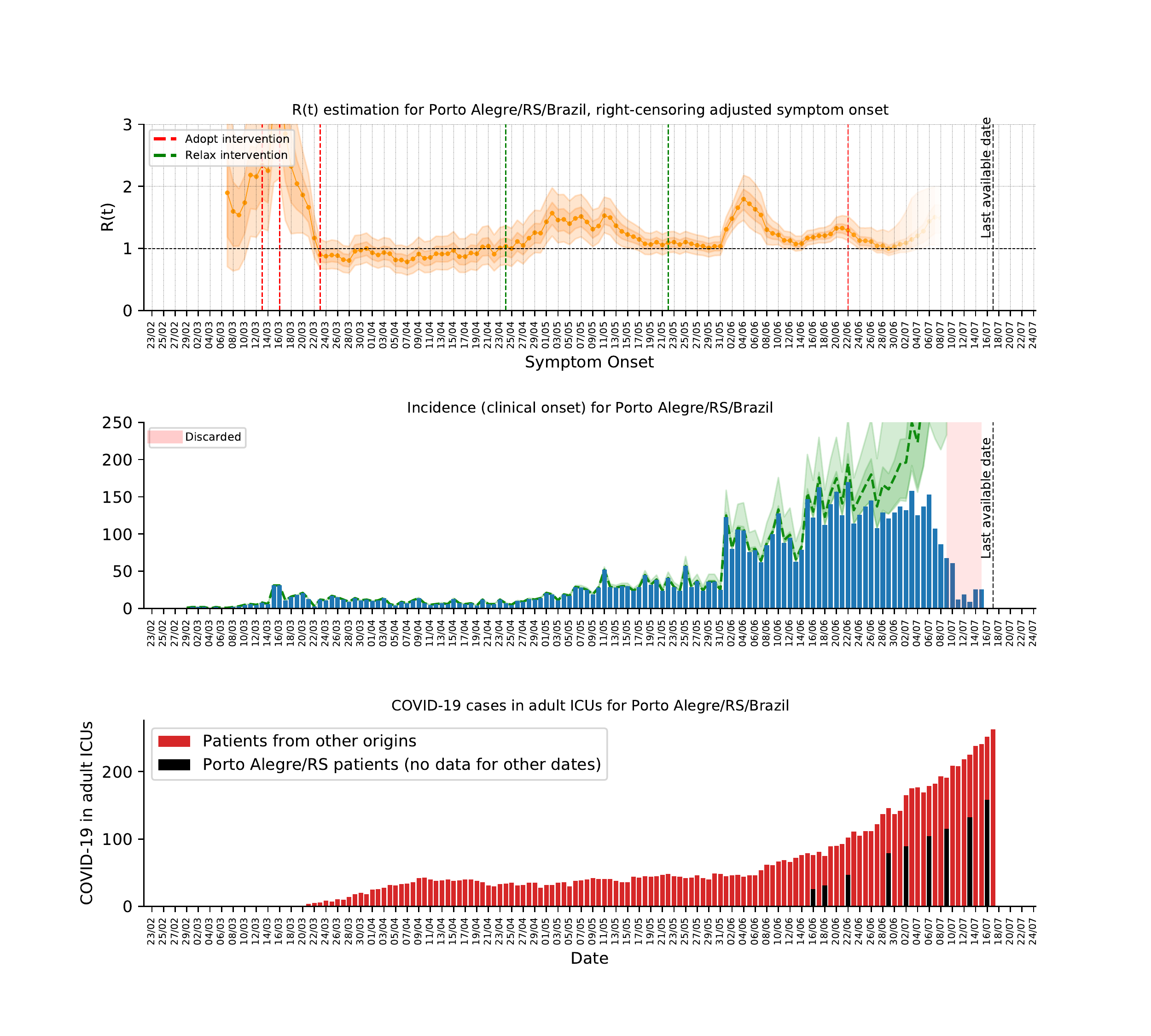}
	\caption{This plot shows the $R_t$ estimation for Porto Alegre/RS. \textbf{Top panel}: in this panel, you can see the $R_t$ estimate with orange bands representing the .5 and .975 posterior HDI, the center line is the median. Some events like adopting interventions or relaxing interventions are marked with green or red vertical dashed lines, respectively. Transparency in the most recent days is used to show uncertainty due to the right-censor adjustment. \textbf{Middle panel}: in the middle panel you can see the incidence by symptom onset in blue and the green dashed line with the bands represent the right-censoring adjustment. The red vertical region represents discarded days as described in the methodology section. \textbf{Bottom panel}: in the bottom panel you can see adult ICU occupancy with confirmed COVID-19 cases. The black bards represent the ICU occupancy by Porto Alegre/RS.}
	\label{fig:rt_poa}
\end{figure}

The same procedure described in the earlier sections was employed to estimate the $R_t$ of Porto Alegre/RS (state capital) during the outbreak. We can see the results of the estimation in Figure \ref{fig:rt_poa}. What we can see in the $R_t$ dynamics is a period of relative stability at the beginning of the outbreak and coincidentally after adopting interventions to reduce mobility. After some time during the restriction period, we were able to see a slow but sustained increase of the $R_t$ and then a bigger increase after relaxing interventions followed by another peak on the $R_t$ potentially due to a sudden change of the testing capacity or reporting procedures.

\subsection{Limitations}
In this analysis, we have not taken into consideration imported cases, which can pose a limitation as it assumes implicitly that all cases after the first, arise from local transmission \cite{Thompson2019}. Note also that interpreting the temporal trends is not always an easy task, changes in $R_t$ can be caused by changes in contact patterns of the affected population, interventions that have their effects overlapped with other interventions, reduction of susceptible population, or changes in test capacity.

\section{ICU length of stay (LoS) and simulation for Porto Alegre/RS}
The state capital of Rio Grande do Sul, Porto Alegre, on July 17th, had an adult ICU occupancy of 263 patients. Among these patients, many were from other cities, however, the major part of them had origin from the city itself (as seen on the black bars at the bottom panel of Figure \ref{fig:rt_poa}).

To estimate the length of stay (LoS) at the ICUs in Porto Alegre/RS we used the data from SIVEP-Gripe\footnote{\url{https://opendatasus.saude.gov.br/dataset/bd-srag-2020}}. While potentially under-reported, this dataset contains a great proportion of the admissions with rich information at the patient level. As shown before on May 18th\footnote{\url{https://perone.github.io/covid19analysis/brazil_icu_stay.html}}, the estimated mean LoS in Rio Grande do Sul was one of the largest in the country.

\begin{figure}[h!]
	\centering
	\begin{minipage}{.6\textwidth}
		\centering
		\includegraphics[width=\linewidth]{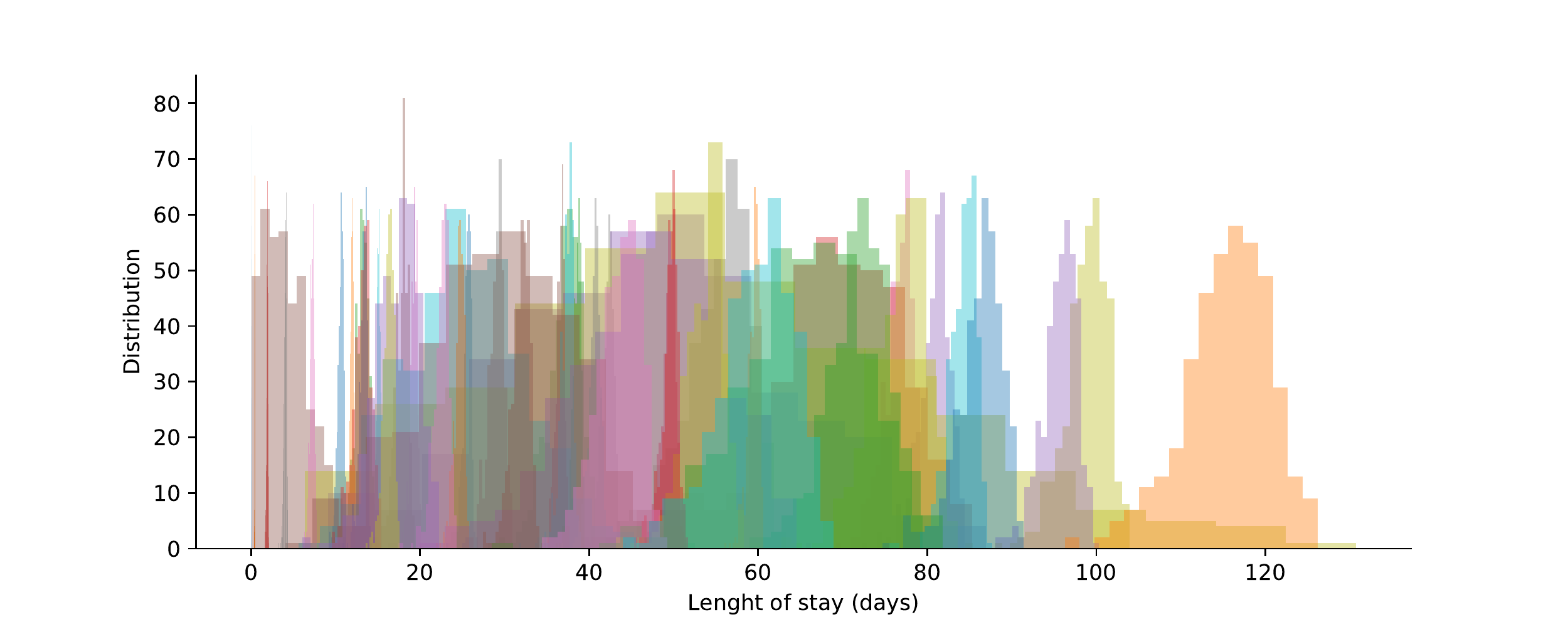}
		\label{fig:test1}
	\end{minipage}%
	\begin{minipage}{.4\textwidth}
		\centering
		\includegraphics[width=\linewidth]{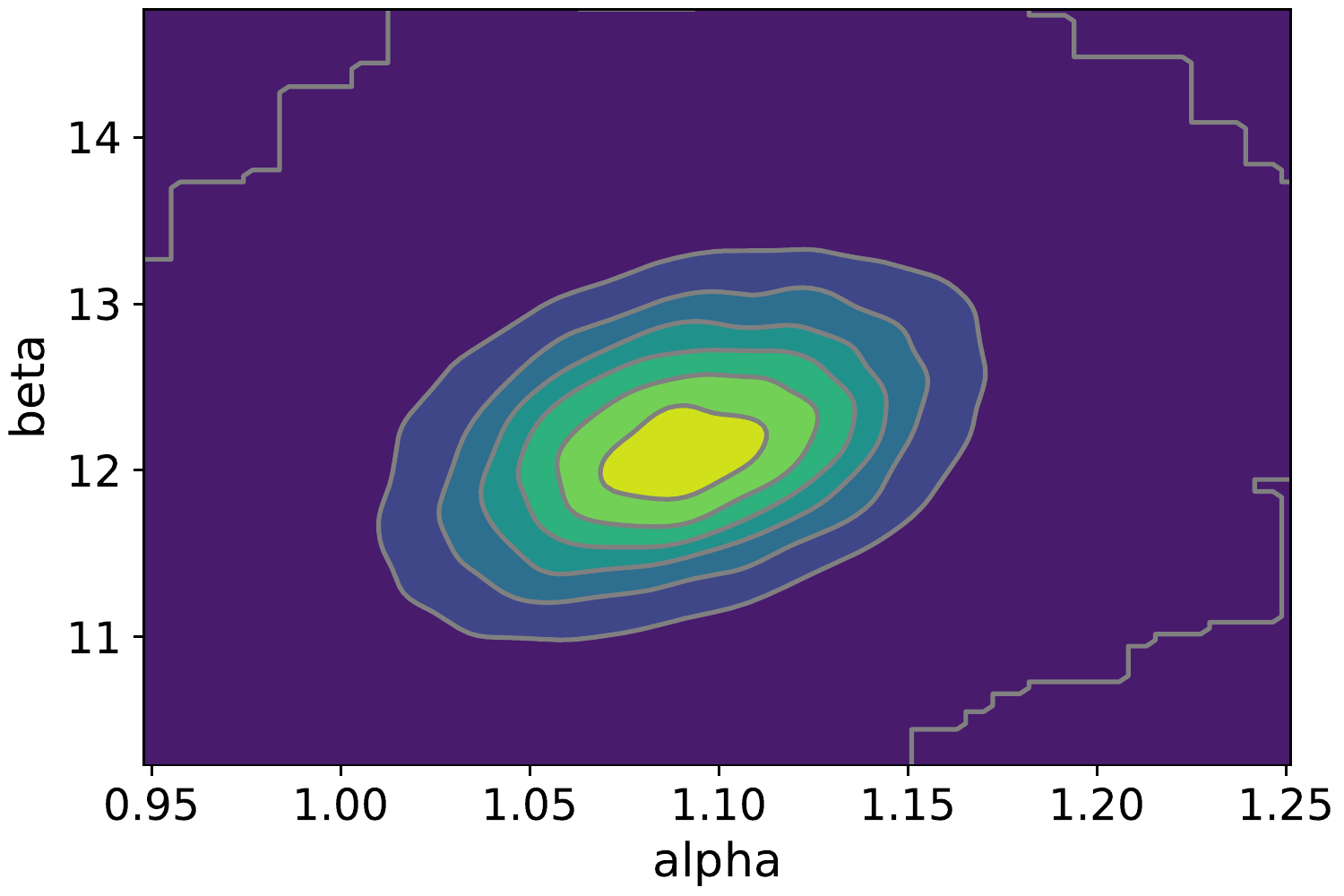}
		\label{fig:test2}
	\end{minipage}
	\caption{\textbf{Left}: Predictive prior distribution for the ICU Length of Stay (LoS) estimation. \textbf{Right}: Estimated posterior distribution for $\alpha$ and $\beta$ of the Weibull distribution for Porto Alegre/RS.}
	\label{fig:los_ppc_posterior}
\end{figure}

To estimate the LoS of Porto Alegre, we did a Bayesian estimation of the parameters $\alpha$ and $\beta$ of a Weibull distribution, using weakly informative priors from a Half Normal distribution with $\sigma=50.0$. Samples from the prior predictive distribution and the estimated posterior distribution can be seen at the left panel and right panel of Figure \ref{fig:los_ppc_posterior}, respectively.

\begin{figure}[h!]
	\centering
	\includegraphics[width=\textwidth]{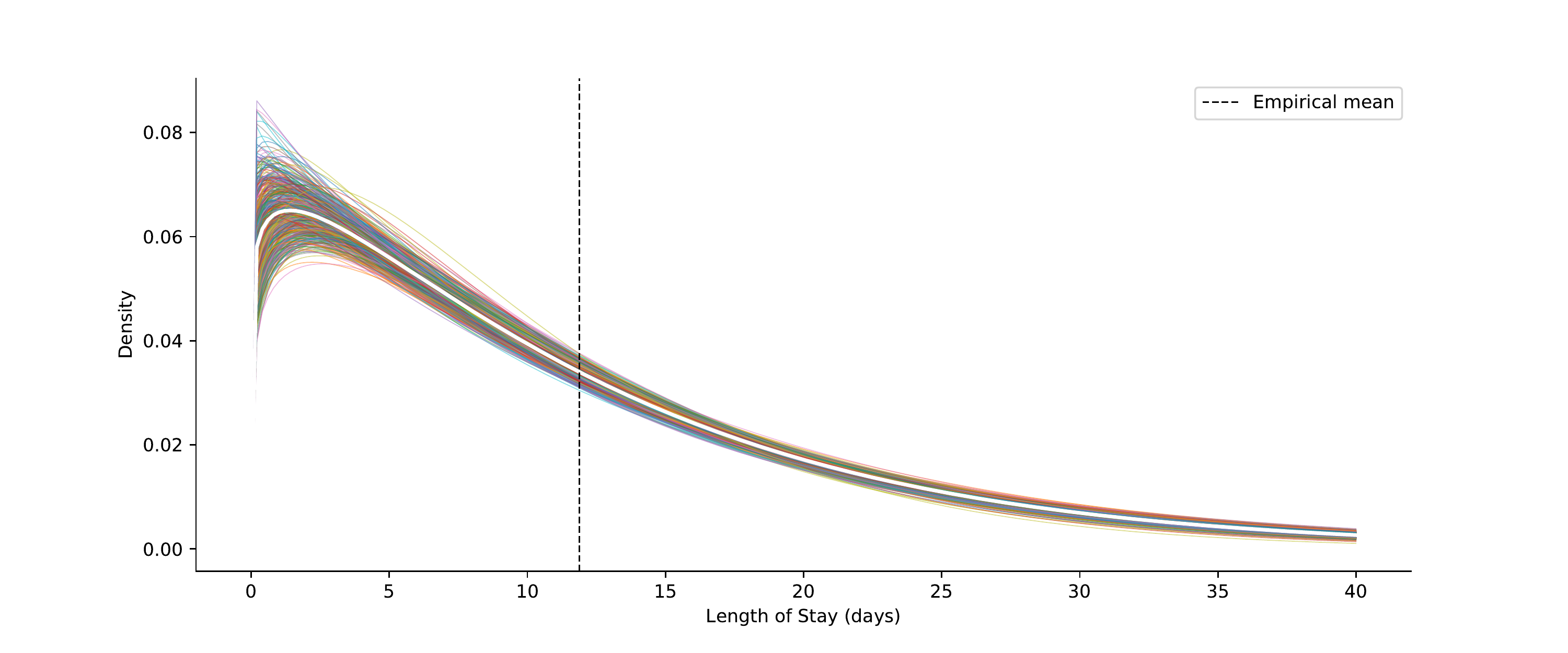}
	\caption{Posterior samples (n=500) from the posterior distribution of the estimated Weibull for the ICU length of stay in Porto Alegre/RS. Truncated at 40 days for visualization purposes. Dashed vertical line is the empirical mean.}
	\label{fig:los_posterior_samples_poa}
\end{figure}

In Figure \ref{fig:los_posterior_samples_poa} we can also see samples from the Weibull posterior distribution. The estimated mean for the $\alpha$ parameter of the Weibull was 1.10 (1.02 - 1.18, HDI 97\%) and 12.42 (11.28 - 13.57, HDI 97\%) for the $\beta$ parameter.

With the estimated LoS distribution at hand, we can now simulate and create future scenarios to understand how ICUs occupation and dynamics will behave. Unfortunately, due to lack of information such as admissions (that are only found at SIVEP-gripe dataset, released in average at every 7 days) and the difficulties to estimate the reporting delays and correct for the right-censoring of this data, we limited the analysis to estimate how long it would take to confidently have a normalization of the ICUs in regarding the COVID-19 patients after stopping the admissions to zero. An unrealistic scenario in the middle of the outbreak, but it shows how long ICU occupancy can take to have the outcome (death/recovery) for all hospitalized patients in these units. We also shown simulated scenarios with different daily admission rate.

Since we are dealing with a discrete quantity (length of stay) in days, we have to discretize the distribution to avoid introducing biases due to rounding, to that purpose we employed \emph{distcrete}\footnote{\url{https://www.repidemicsconsortium.org/distcrete/}}. We also discarded the last 7 days of data due to right-censoring.

To simulate the ICU we used a similar approach described in \cite{JombartThibautandNightingaleEmilySandJitMarkanddeWarouxOandKnightGwenandFlascheStefanandEggoRosalindandKucharskiAdamJandPearsonCarlABandProcter}, where for each date and incidence of admissions on that day we sample an estimate from the LoS posterior and then beds are counted by summing all cases for each day. To take stochasticity of the LoS into account, this process is repeated N times, resulting in N predictions of bed needs over time, where the credibility intervals are computed.

\begin{figure}[h!]
	\centering
	\includegraphics[width=\textwidth]{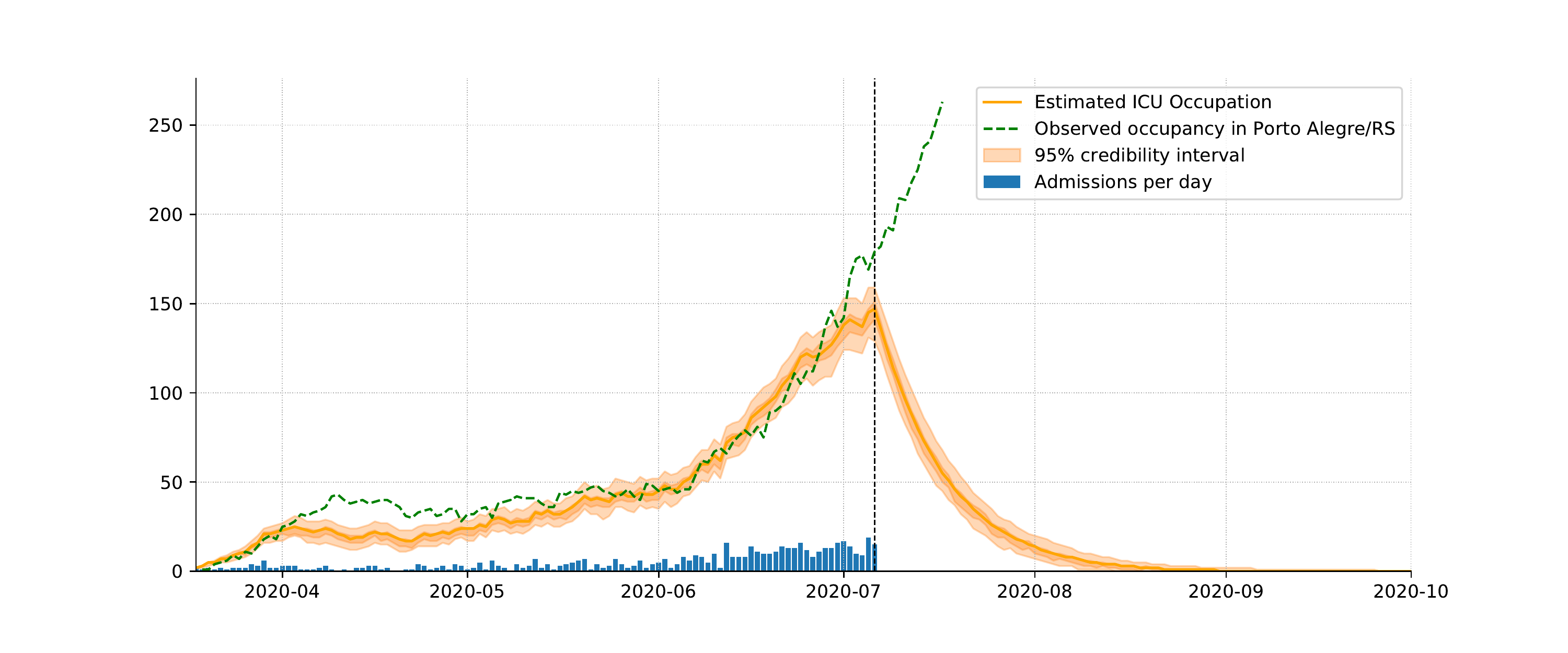}
	\caption{ICU occupation simulation with estimated LoS (length of stay) distribution from Porto Alegre/RS. Admissions per day are based on SIVEP-Gripe data provided by Brazilian Ministry of Health, Simulation was done using 500 rounds. Orange bands are .5 and .95 HDI uncertainty intervals. Last seven days of the SIVEP-Gripe dataset were discarded due to right-censoring, the last available admission date is on July 6th, 2020.}
	\label{fig:icu_simulation_poa}
\end{figure}

The result of this simulation can be seen in Figure \ref{fig:icu_simulation_poa}. As we can see, even with initial under-reporting of SIVEP-Gripe data and right-censoring, we were able to match the complex dynamics of the daily observed ICU occupancy in Porto Alegre/RS. Note also that the simulation was able to match the growing observed ICU occupancy very well even without using the release dates from SIVEP-Gripe, showing that the model is indeed capturing the observed growing patterns in Porto Alegre/RS. What is worrying about this plot is that if admissions due to COVID-19 are stopped to \emph{zero}, it would \emph{take nearly one month and a half to decay the occupancy to zero}.

\begin{figure}[h!]
	\centering
	\includegraphics[width=\textwidth]{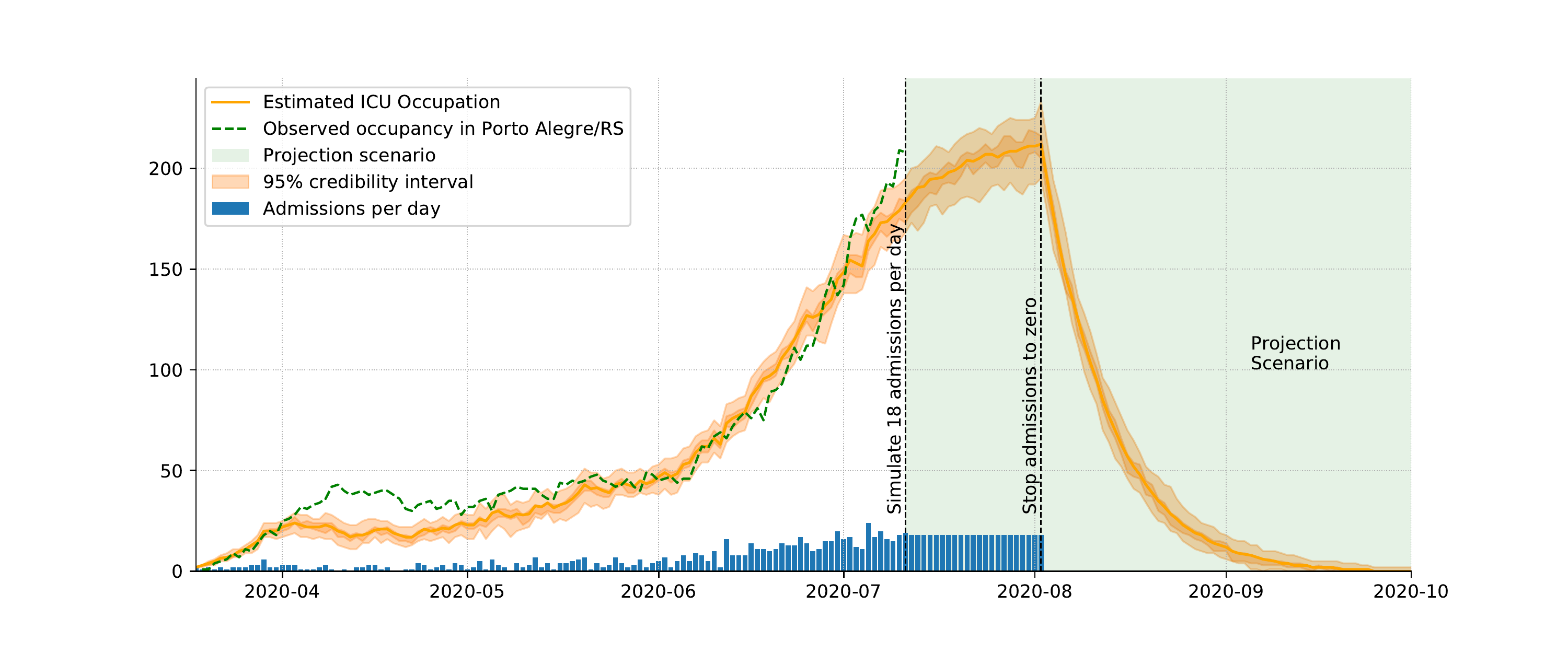}
	\includegraphics[width=\textwidth]{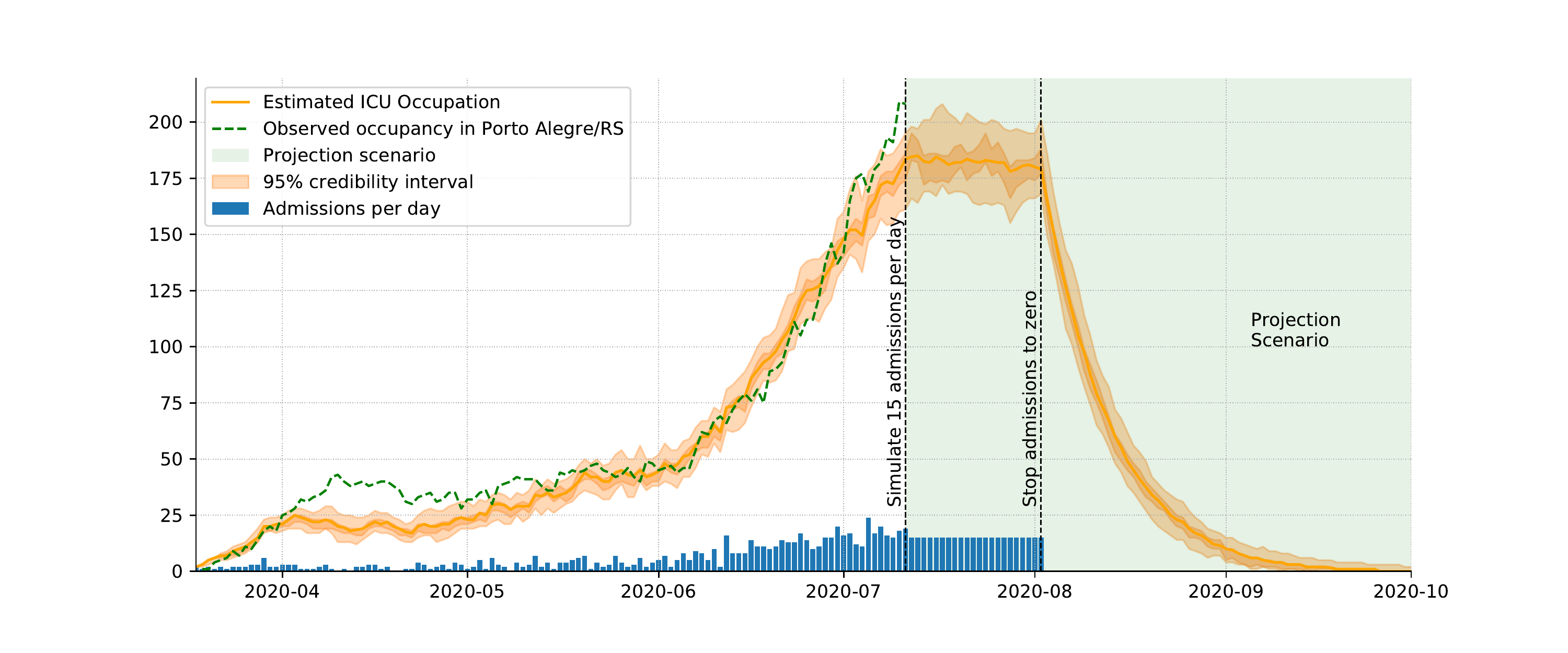}
	\includegraphics[width=\textwidth]{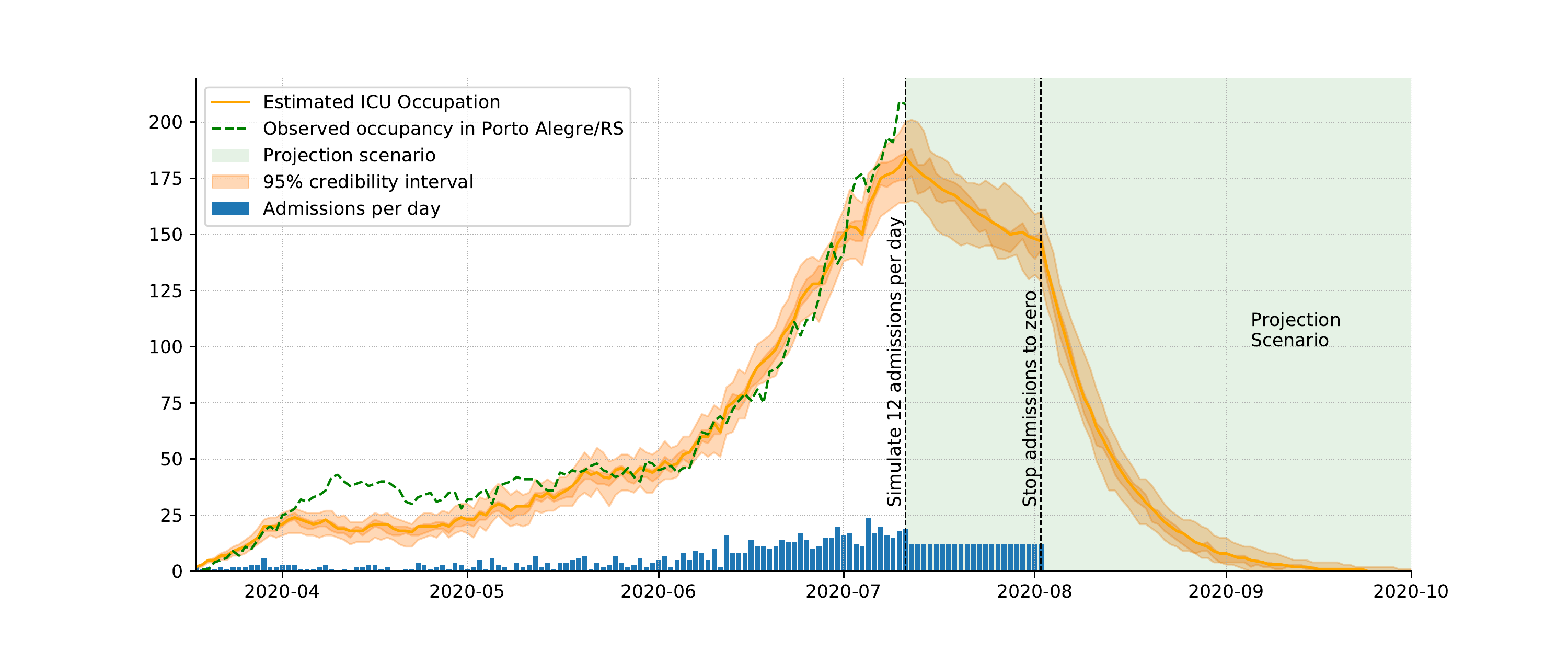}
	
	\caption{ICU occupation simulation scenarios with estimated LoS (length of stay) distribution from Porto Alegre/RS. \textbf{Top:} scenario where we simulate a constant rate of 18 daily admissions for a period of time. \textbf{Middle:} scenario with a simulated constant rate of 15 daily admissions. \textbf{Bottom:} scenario with a simulated constant rate of 12 daily admissions. Admissions per day are based on SIVEP-Gripe data provided by Brazilian Ministry of Health and are shown until the beginning of the constant rate simulation. Simulation was done using 300 rounds. Orange bands are .5 and .95 HDI uncertainty intervals. Green shade represents simulation for the different scenarios.}
	\label{fig:icu_projection_poa}
\end{figure}

In Figure \ref{fig:icu_projection_poa}, we show a simulation for three different scenarios using a period with constant daily admission rates of 12, 15, and 18 before dropping admissions to zero. As we can see in the figures, there is evidence to support that a daily admission rate below 15 patients would be required to start to see a downward trend in the occupation of the ICUs in Porto Alegre/RS. Note that in the figures, there is a period where we use reported admissions from SIVEP-Gripe data, then followed by a constant daily admission rate for a period and finally we stop admissions (to zero). These three different periods are to show how ICU occupancy will behave under different scenarios. The different scenarios in Figure \ref{fig:icu_projection_poa} show how fast occupancy can grow, but how slow it is to decay, as seen on many countries suffering from high ICU occupancy.

\begin{figure}[h!]
	\centering
	\includegraphics[width=1.2\textwidth]{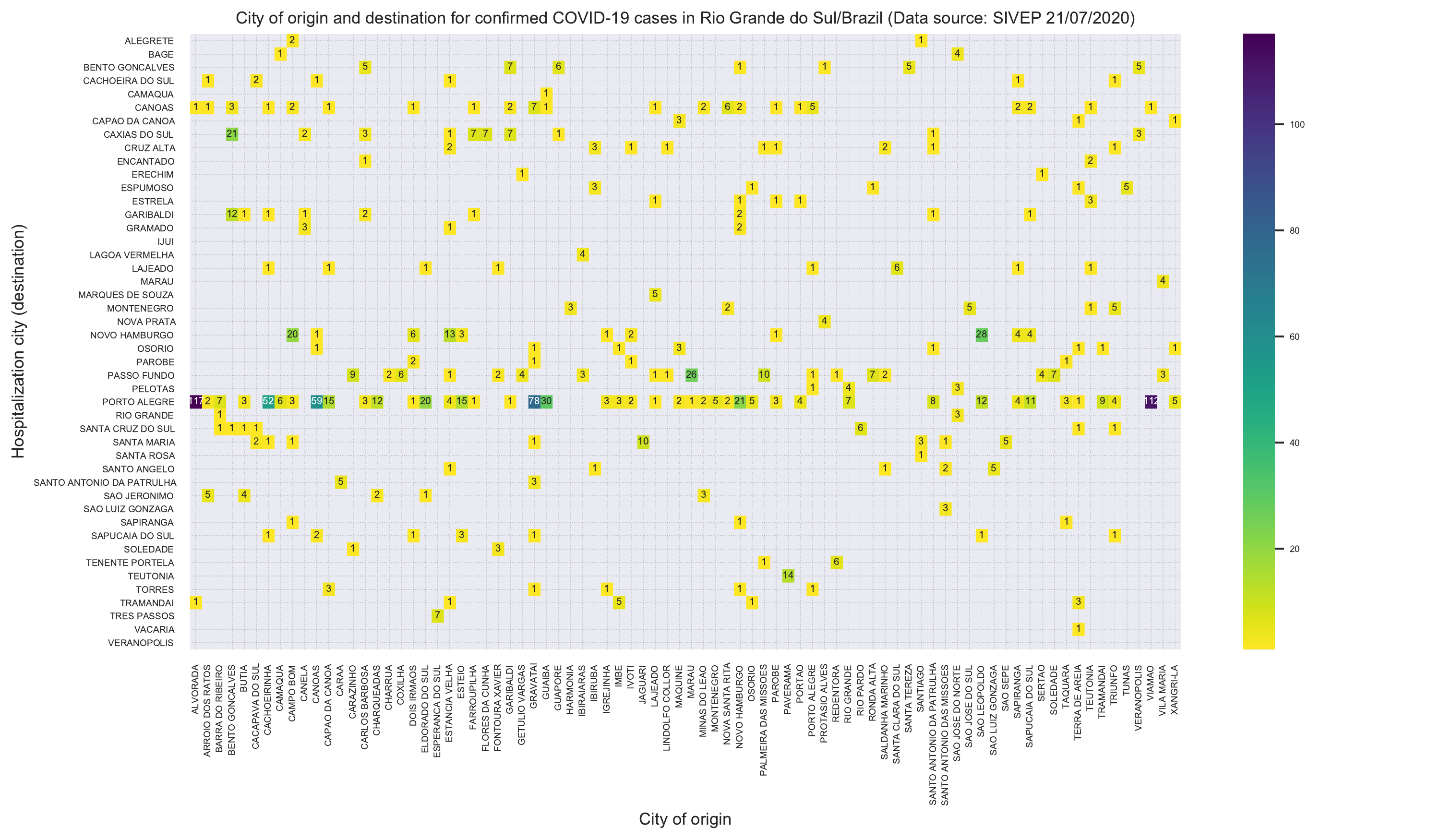}
	\caption{This graph shows the number of hospitalizations found on SIVEP-Gripe dataset where the hospitalization city is different then the city of residence. The plot is constrained to cities where the sum of all desitination or sources is greater or equal then 5 patients. The number at each square is the number of hospitalizations with the color map mapping the range of hospitalizations. This graph used data from SIVEP-Gripe dated as 21th July, 2020.}
	\label{fig:hospitalizations_rs}
\end{figure}

\begin{figure}[h!]
	\centering
	\includegraphics[width=1.1\textwidth]{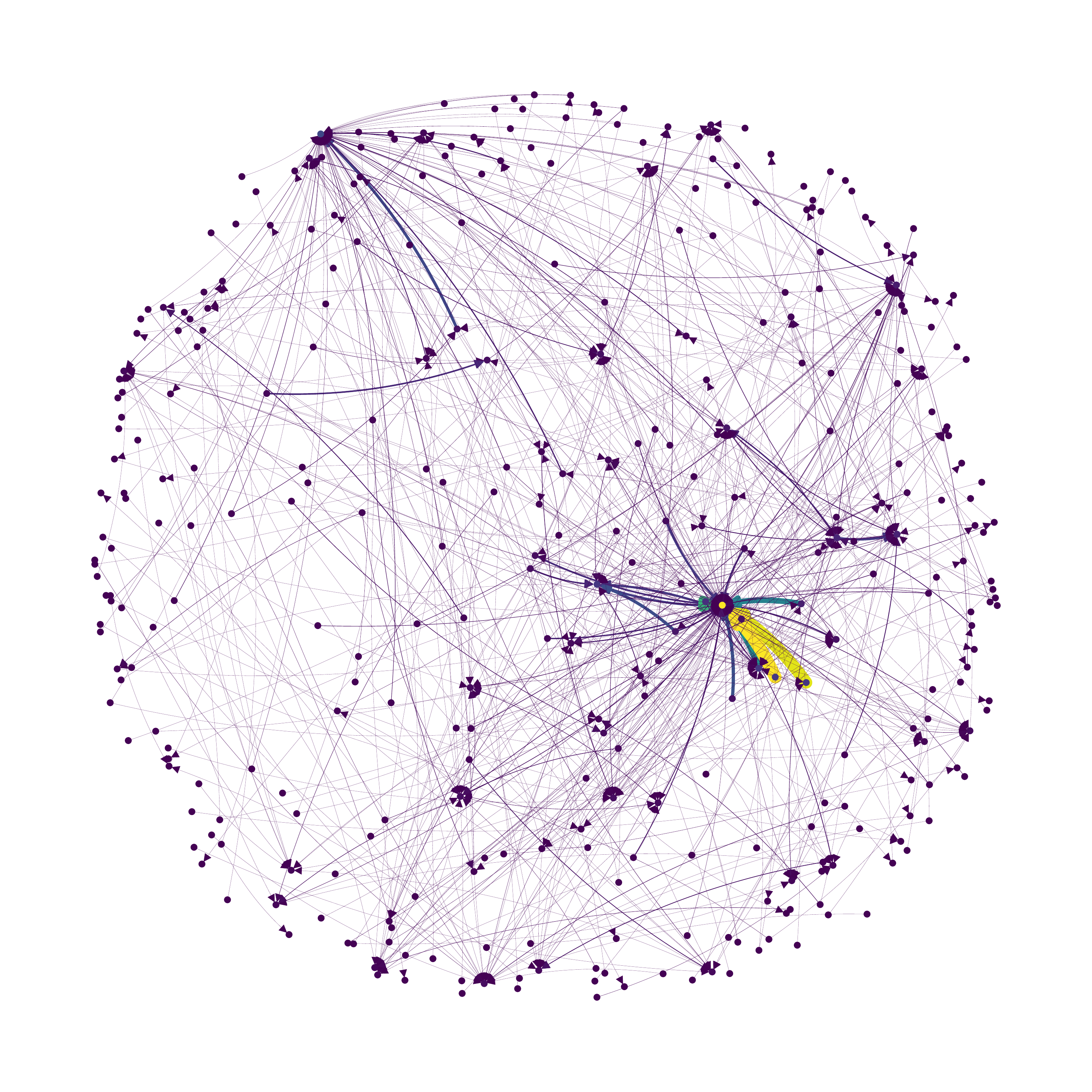}
	\caption{This is a directed graph plot showing the complexity of hospitalizations in Rio Grande do Sul/Brazil. Each node represents a city and the edges shows the direction of hospitalization from patients on the city of origin to the destination city. Colors and thickness of the edge represents the amount of hospitalizations from a city to another (the lighter the color or thicker the edge, the more hospitalizations had origin at the city). The node with yellowish arrows arriving at close to the center of the graph is Porto Alegre / RS. This graph used data from SIVEP-Gripe dated as 21th July, 2020.}
	\label{fig:hosp_rs_graph}
\end{figure}

\begin{figure}[h!]
	\centering
	\includegraphics[width=1.1\textwidth]{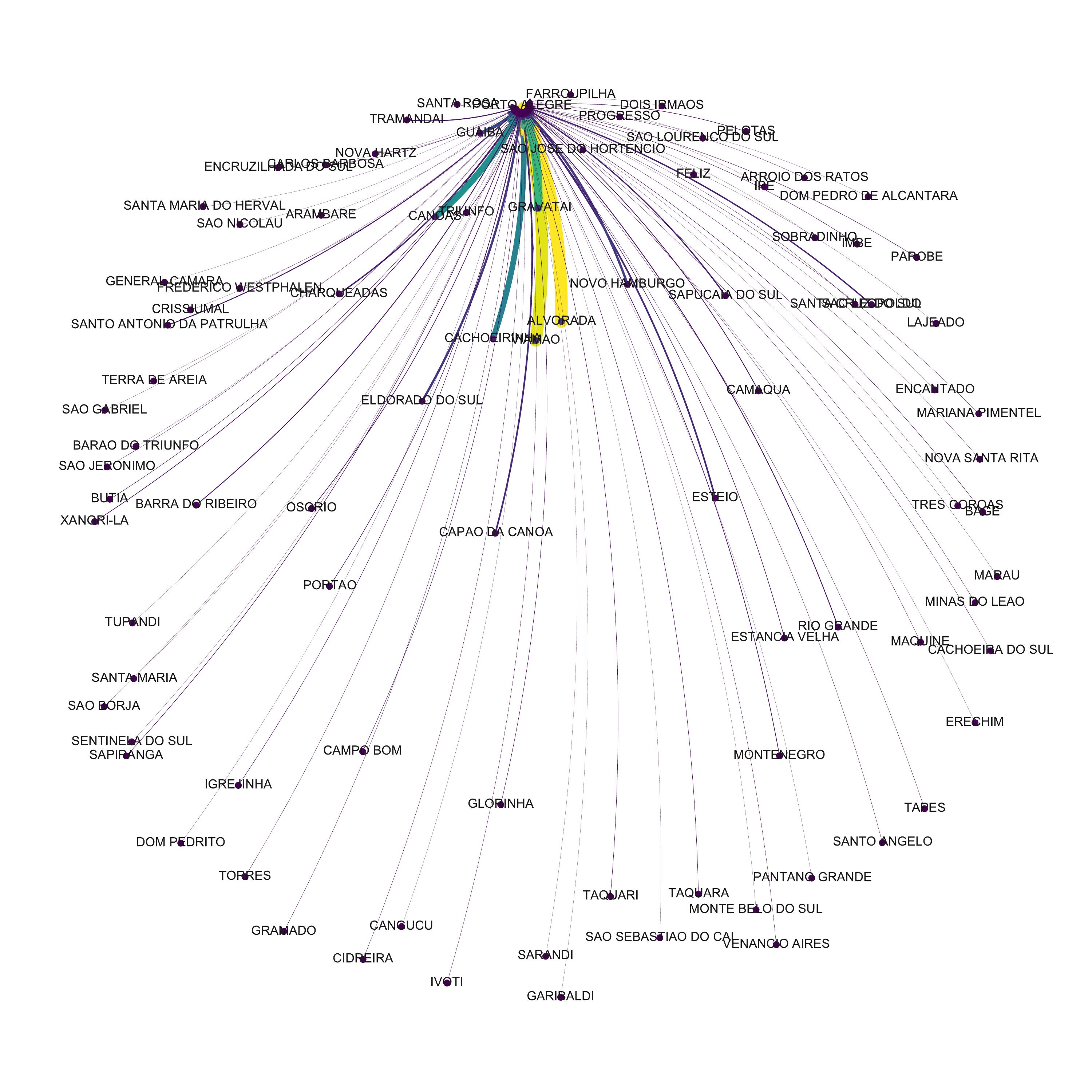}
	\caption{This is a directed graph plot showing the city of residence of patients hospitalized in Porto Alegre / RS. Each node represents a city and the edges shows the direction of hospitalization from patients on the city of origin to Porto Alegre/RS. Colors and thickness of the edge represents the amount of hospitalizations from a city to another (the lighter the color or thicker the edge, the more hospitalizations had origin at the city). This graph used data from SIVEP-Gripe dated as 21th July, 2020.}
	\label{fig:graph_hosp_poa}
\end{figure}

\begin{figure}[h!]
	\centering
	\includegraphics[width=1.1\textwidth]{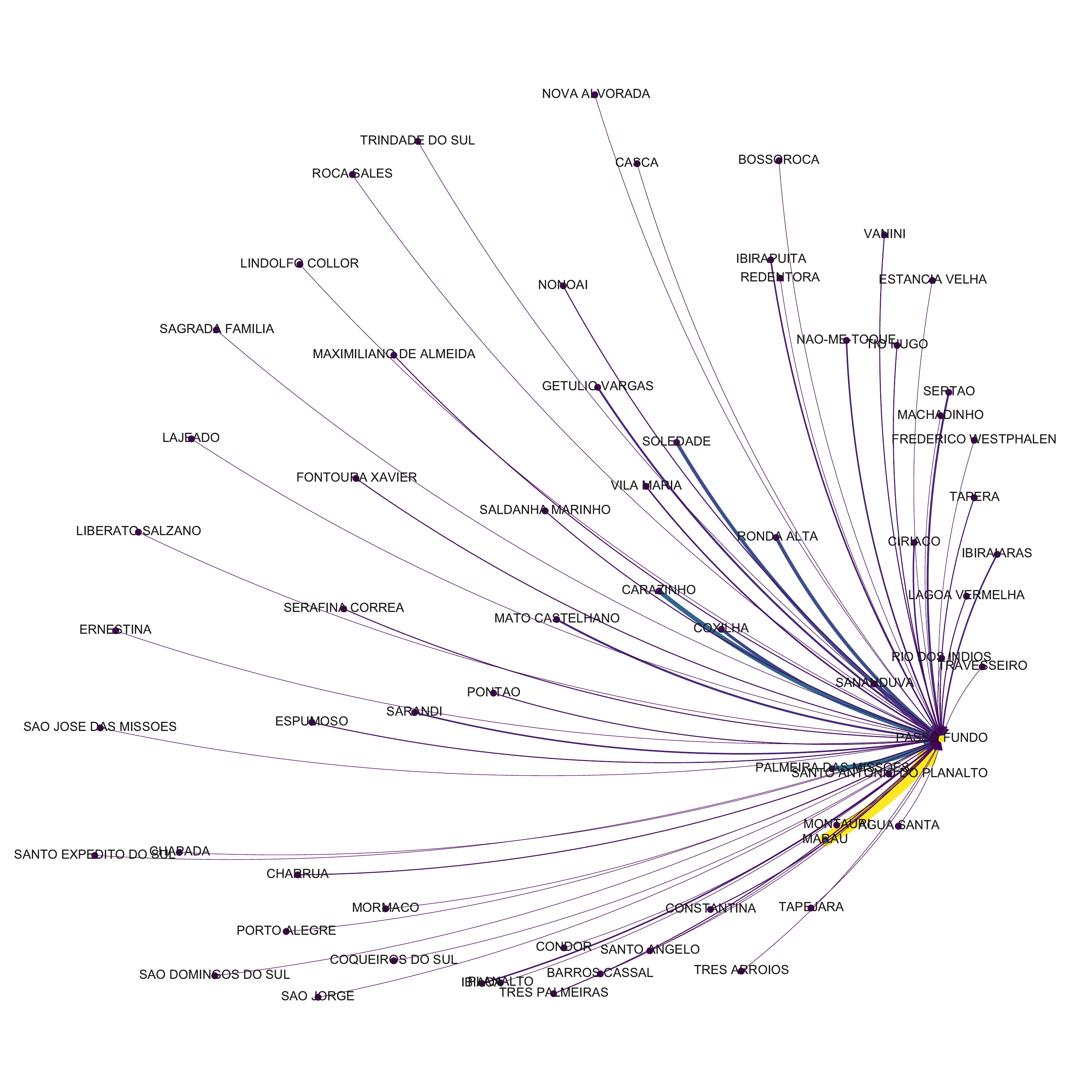}
	\caption{This is a directed graph plot showing the city of residence of patients hospitalized in Passo Fundo / RS. Each node represents a city and the edges shows the direction of hospitalization from patients on the city of origin to Passo Fundo/RS. Colors and thickness of the edge represents the amount of hospitalizations from a city to another (the lighter the color or thicker the edge, the more hospitalizations had origin at the city). This graph used data from SIVEP-Gripe dated as 21st July, 2020.}
	\label{fig:graph_hosp_pf}
\end{figure}

Regarding general hospitalizations, we show in Figure \ref{fig:hospitalizations_rs} a table with hospitalizations present in the SIVEP-Gripe dataset where the hospitalization city is different than the city of residence. The plot is constrained to cities where the sum of all desitination or sources is greater or equal than 5 patients. We can see the enormous pressure that neighbour cities can cause to health hubs such as Porto Alegre/RS and Passo Fundo/RS. We show the complexity of these import/export dynamics in the Figure \ref{fig:hosp_rs_graph} where we plot a directed graph for all cities in Rio Grande do Sul. Finally, in Figure \ref{fig:graph_hosp_poa} and Figure \ref{fig:graph_hosp_pf} we show the same directed graph but only when hospitalization city is Porto Alegre/RS and Passo Fundo/RS, respectively.

\subsection{Limitations}
For this analysis we assume that SIVEP-Gripe provides a representative report of the ICU hospitalizations in Porto Alegre, however, it might be biased and lack reports from some health institutions. Admission dates can also be lagged, therefore it can introduce bias in the analysis as well. However, by looking at the simulation results we can see that it approximated the observed daily aggregated reports from all hospitals of the city with data provided from the city ICU occupancy dashboard \footnote{\url{https://bit.ly/3hkXXmd}}.

\section{City-level mobility analysis}
Mobility data from providers such as Google~\cite{Google2020} and Facebook~\cite{Facebook} that are collected from mobile devices provide us a unique view on mobility. Although mobility data from Google can help at the state-level analyses, by leveraging higher resolution data such as mobility data from Facebook~\cite{Facebook}, we can have rich information about the effect of adopted interventions at the city level.

\begin{figure}[h!]
	\centering
	\includegraphics[width=\textwidth]{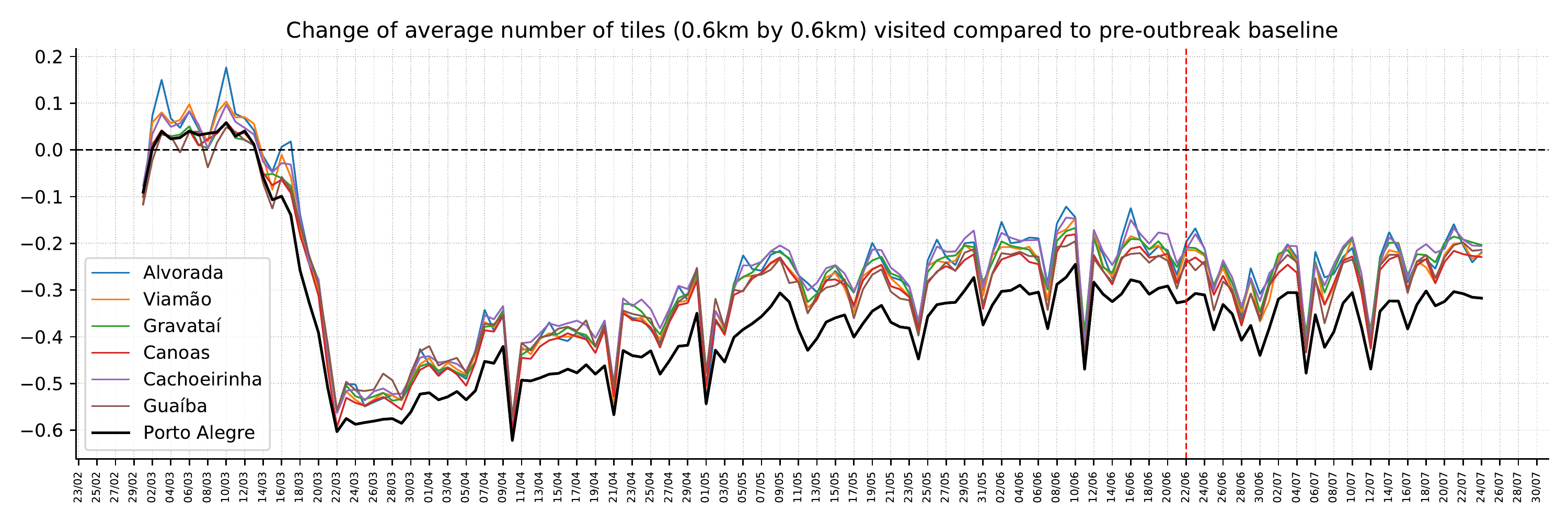}
	\includegraphics[width=\textwidth]{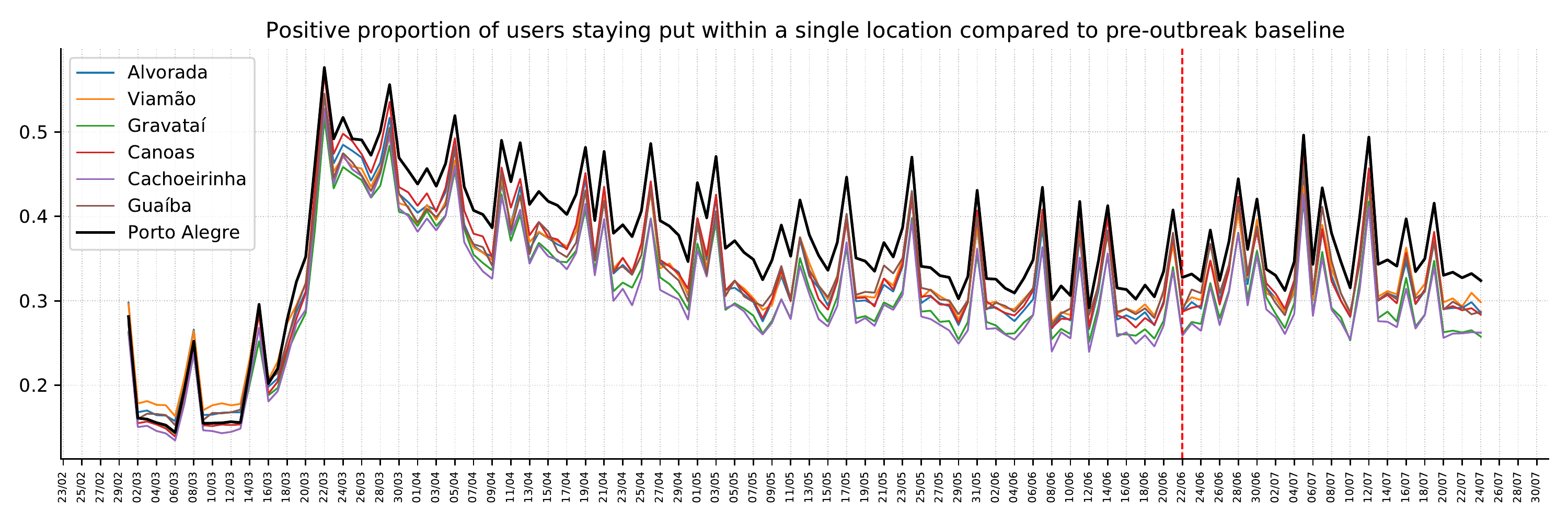}
	
	\caption{This figure shows mobility data for Porto Alegre/RS (in black, thicker line), together with the mobility of the 6 cities that are exporting hospitalizations to Porto Alegre/RS (Alvorada, Viamão, Gravataí, Canoas, Cachoeirinha, Guaíba). The dashed red lines shows the date when the mayor of Porto Alegre/RS released enactment with new interventions. \textbf{Top:} in this panel we show the change of average number of tiles (0.6km by 0.6km) visied, clipped to 200 maximum. \textbf{Bottom:} the positive proportion of users staying put within a single location, it represents people who are only observed in a single level tile during the course of a day. This plot used hospitalizations from SIVEP-Gripe dataset dated of 21st July, 2020. It also used mobility data from Facebook \cite{Facebook}.}
	\label{fig:mobility_change}
\end{figure}

In Figure~\ref{fig:mobility_change}, we show mobility data for Porto Alegre/RS, together with the mobility of the top 6 cities that are exporting hospitalizations to Porto Alegre/RS, to mention: Alvorada, Viamão, Gravataí, Canoas, Cachoeirinha, Guaíba. In this figure, we can see that after the initial peak of interventions around the last week of March, there was a slow increase in the change of average number of visited tiles (represented by 0.6km by 0.6km) compared to the pre-outbreak baseline. After new interventions on July 22th, we can see reduction again of mobility, but not to the same levels seen before. We can also see in this figure, that neighbour cities (that are exporting hospitalizations to Porto Alegre/RS) doesn't seem to have the same level of agreement with the mobility reduction seen in Porto Alegre/RS, especially when we look at the weekends. Also, looking at the weekend of July 19th, when compared to the weekend of 12th July, we can see an increase of mobility not only in Porto Alegre/RS but also in the cities exporting hospitalizations to the city, a worrying scenario as ICUs in Porto Alegre/RS reached already nearly 90\% of occupancy at these dates.

\begin{figure}[h!]
	\centering
	\includegraphics[width=\textwidth]{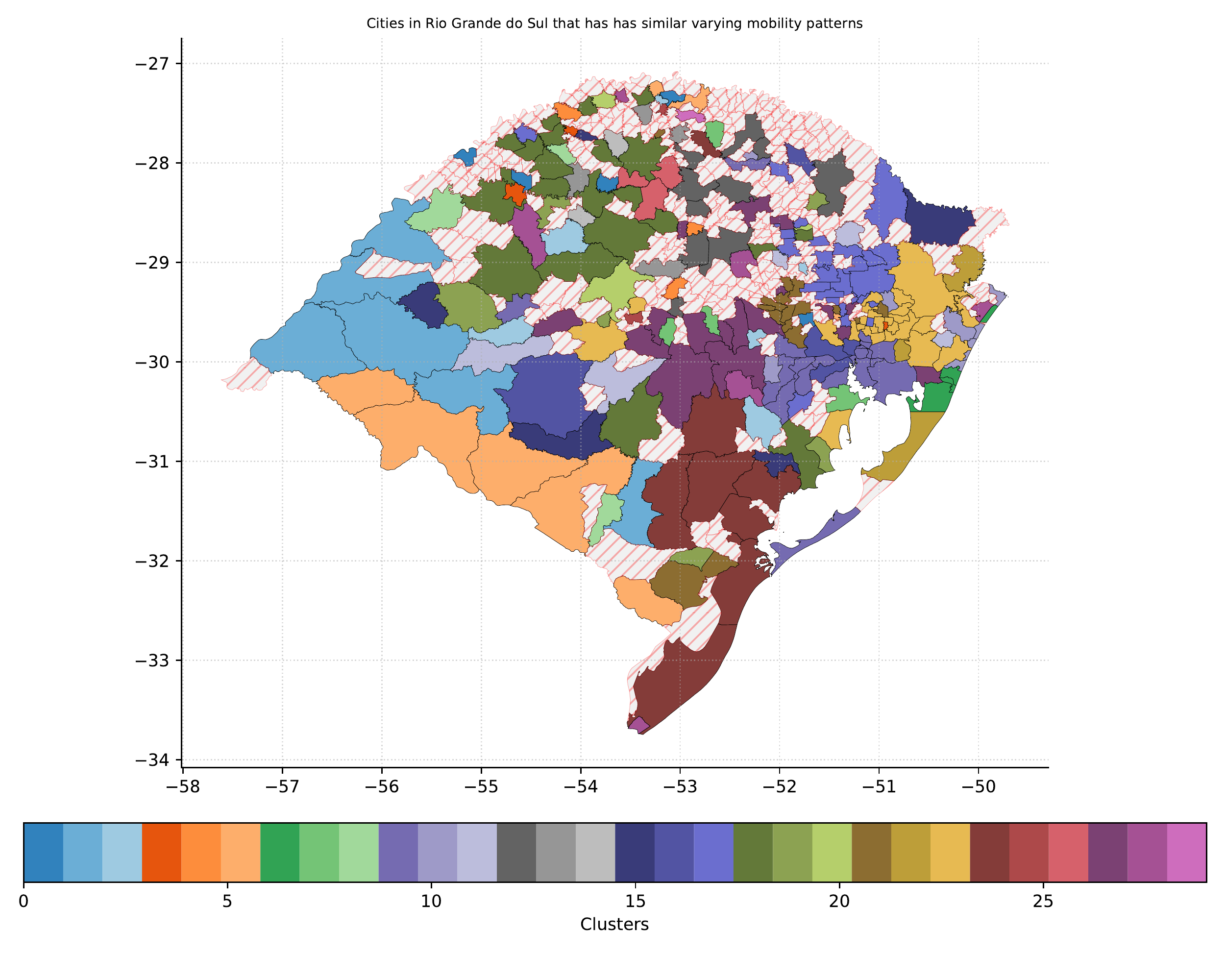}
	
	\caption{This map shows the mobility clusters found using affinity propagation on the estimated sparse inverse covariance from mobility patterns. Cities with similar patterns are in the same cluster (same color). This map used mobility data from Facebook \cite{Facebook}. Regions with parallel red lines are regions with missing mobility data or partial mobility data that wasn't included in the analysis.}
	\label{fig:rs_mob_clusters}
\end{figure}

\begin{figure}[h!]
	\centering
	\includegraphics[width=\textwidth]{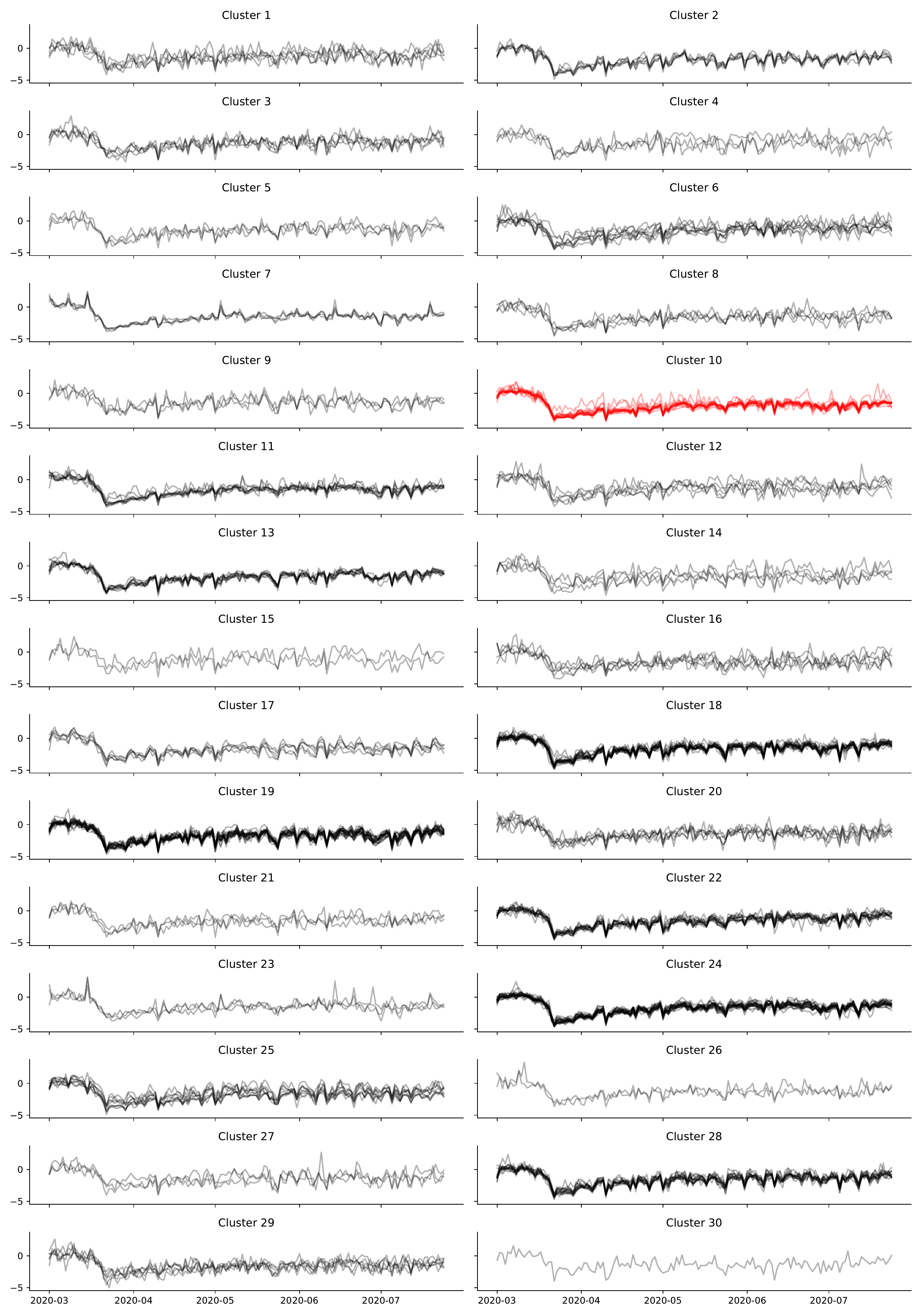}
	\caption{Clusters of mobility after affinity propagation on the estimated sparse inverse covariance (precision matrix). Note that the the series are all standardized. The cluster in red is to show thich cluster Porto Alegre/RS belongs to.}
	\label{fig:mobility_clusters}
\end{figure}

Another interesting analysis that can be done with city-level mobility data, is to find what are the cities that share similar mobility patterns. We use sparse inverse covariance (precision matrix) estimation to find which cities are correlated conditionally on the others using mobility data. From the precision matrix, we employ affinity propagation~\cite{Frey2007}, a clustering technique that creates clusters by sending messages between pairs of samples until convergence is achieved. Please note that we are interested in patterns that vary together, therefore we standardize mobility patterns. Affinity propagation found 30 clusters of cities that are described in Table \ref{table:mob_clusters_table}. We also show the mobility patterns beloging to each cluster in Figure \ref{fig:mobility_clusters} and finally a geographical map of Rio Grande do Sul with colors representing the clusters that each city belongs to in Figure \ref{fig:rs_mob_clusters}.

\begin{table}[bt]
	\small
	\caption{Table showing the 30 clusters found using affinity propagation on the estimated sparse inverse covariance of the Facebook's mobility data~\cite{Facebook}.}
	\begin{threeparttable}
		%\begin{tabularx}{lccrr}
		\begin{tabularx}{\textwidth}{|l|X|}
			\headrow
			\thead{Cluster} & \thead{Cities} \\
			Cluster 1 & Ajuricaba, Guarani das Missões, Paverama, Planalto, Porto Xavier \\
			Cluster 2 & Alegrete, Itaqui, Pinheiro Machado, Rosário do Sul, São Borja, Uruguaiana \\
			Cluster 3 & Ametista do Sul, Dom Feliciano, Jóia, Passo do Sobrado, São Valentim do Sul \\
			Cluster 4 & Araricá, Boa Vista do Buricá, Caibaté \\
			Cluster 5 & Arroio do Tigre, Crissiumal, Tapera \\
			Cluster 6 & Baje, Dom Pedrito, Harmonia, Iraí, Jaguarão, Nonoai, Quaraí, Santana do Livramento \\
			Cluster 7 & Arroio do Sal, Balneário Pinhal, Cidreira, Palmares do Sul \\
			Cluster 8 & Barra do Ribeiro, Campinas do Sul, Paraíso do Sul, Vale do Sol \\
			Cluster 9 & Candiota, Independência, Santo Antônio das Missões \\
			Cluster 10 & Alvorada, Arroio dos Ratos, Butiá, Cachoeirinha, Canoas, Charqueadas, Esteio, General Câmara, Gravataí, Guaíba, Jaguari, Porto Alegre, São Jerônimo, São Leopoldo, Sapucaia do Sul, Sertão, Viamão \\
			Cluster 11 & Capão do Leão, Estância Velha, Gramado, Imbé, Maquiné, Minas do Leão, Torres, Tramandaí, Xangri-lá \\
			Cluster 12 & Anta Gorda, Cacequi, Caraá, Ipê, São Sepé \\
			Cluster 13 & Campo Real, Carazinho, Erechim, Espumoso, Getúlio Vargas, Lagoa Vermelha, Passo Fundo, Sarandi, Sobradinho, Soledade \\
			Cluster 14 & Catuípe, Constantina, Salto do Jacuí, Seberi \\
			Cluster 15 & Augusto Pestana, Coronel Bicaco \\
			Cluster 16 & Bom Jesus, Cristal, Lavras do Sul, Manoel Viana, São Martinho \\
			Cluster 17 & Eldorado do Sul, Nova Santa Rita, Sananduva, São Gabriel, Triunfo \\
			Cluster 18 & Antônio Prado, Baro, Bento Gonçalves, Bom Princípio, Carlos Barbosa, Caxias do Sul, Dois Irmãos, Farroupilha, Flores da Cunha, Garibaldi, Guaporé, Nova Petrópolis, Paraí, São Marcos, São Sebastião do Caí, Serafina Corrêa, Tapejara, Tuparendi, Vacaria, Vale Real, Veranópolis \\
			Cluster 19 & Arvorezinha, Caçapava do Sul, Camagua, Cerro Largo, Cruz Alta, Frederico Westphalen, Giruá, Horizontina, Ibirubá, Ijuí, Nova Bassano, Palmeira das Missões, Panambi, Santa Rosa, Santiago, Santo Ángelo, Santo Augusto, Santo Cristo, São Luiz Gonzaga, Três de Maio, Três Passos, Tupanciretã \\
			Cluster 20 & Arambaré, Entre-Ijuís, Ibiraiaras, Itaara, Pedro Osório, São Francisco de Assis \\
			Cluster 21 & Júlio de Castilhos, Nova Araçá, Tenente Portela \\
			Cluster 22 & Arroio do Meio, Arroio Grande, Bom Retiro do Sul, Cruzeiro do Sul, Encantado, Estrela, Lajedão, Lindolfo Collor, Roca Sales, Santa Clara do Sul, Santa Maria do Herval, Taquari, Teutônia, Tupandi \\
			Cluster 23 & Cambará do Sul, Glorinha, Mostardas \\
			Cluster 24 & Campo Bom, Canela, Estação, Igrejinha, Ivoti, Montenegro, Morro Reuter, Nova Hartz, Nova Palma, Novo Hamburgo, Osório, Parobé, Picada Café, Portao, Rolante, Santa Maria, Santo Antônio da Patrulha, São Francisco de Paula, Sapiranga, Tapes, Taquara, Terra de Areia, Três Coroas \\
			Cluster 25 & Canguçu, Capitão, Encruzilhada do Sul, Pelotas, Piratini, Rio Grande, Ronda Alta, Santa Vitória do Palmar, São Lourenço do Sul \\
			Cluster 26 & Faxinal do Soturno, Rodeio Bonito \\
			Cluster 27 & Chapada, Condor, Santa Bárbara do Sul \\
			Cluster 28 & Agudo, Cachoeira do Sul, Candelária, Capão da Canoa, Casca, Feliz, Maraú, Restinga Seca, Rio Pardo, Salvador do Sul, Santa Cruz do Sul, São Pedro do Sul, Selbach, Venâncio Aires, Vera Cruz \\
			Cluster 29 & Chuí, Fontoura Xavier, Palmitinhos, Pantano Grande, São Miguel das Misses, Três Cachoeiras \\
			Cluster 30 & Trindade do Sul \\
			\hline  % Please only put a hline at the end of the table
		\end{tabularx}
	\end{threeparttable}
	\label{table:mob_clusters_table}
\end{table}

\subsection{Limitations}
Interpreting mobility data is challenging. The information from providers on population mobility, needs to be contextualized with additional data sources on local demographics, infrastructure and socioeconomic indicators. As mentioned in~\cite{Foundation2020}, areas with higher densities of population and infrastructure, tend to have shorter trips and easier access to essential services, while the opposite can happen on other regions. Also, data from Facebook mobility represent only a slice of users that have opted into the Location History setting, therefore it may be overrepresented in certain populations and underrepresented in others.

\section{Discussion}
Aware of all limitations and difficulties to estimate and interpret the $R_t$ through time, we provided an adjusted estimation of this important epidemiological quantity that is a fundamental variable to observe during the outbreak and within the context. We also provided an estimation for the length of stay (LoS) for the state capital Porto Alegre, where we showed how fast admissions can cause a surge of occupancy and how slow are these occupancies to decay to the same regime it was before the surge.

The goal of this work was not only to try to characterize the dynamics of the outbreak in the state but also to raise awareness to the population to the worrying scenario that usually comes after a long period when even basic epidemiological quantities such as $R_t$ are above their safe regimes. While it is true that there are many limitations in these estimates, when we look at these quantities in context together with mobility data and ICU progression, we can see that urgent interventions are needed to break the chain of transmissions and the growing trend hospitalizations that we are witnessing for more than 2 months.

\clearpage

\section*{acknowledgements}
We are very thankful for the work done by Brasil.io and its contributors. We are also thankful to Fernando Azambuja for the help with the SES-RS historical files. We are also thankful to everyone from Infovid\footnote{\url{https://twitter.com/grupo_infovid}} and Rede Análise COVID-19\footnote{\url{https://twitter.com/analise_covid19}} for all the help and fruitful discussions.

\bibliography{library}

\end{document}